\documentclass[%
preprint,
nofootinbib,
 amsmath,amssymb,
 aps,
]{revtex4-2}

\usepackage{graphicx}% Include figure files
\usepackage{dcolumn}% Align table columns on decimal point
\usepackage{bm}% bold math
\usepackage{hyperref}% add hypertext capabilities
\begin{document}
%\title{Mechanical binding and self-assembled chains of granular particles with rotational activity}
\title{Mechanical pairing and self-assembled chains of non-convex granular rotors}
\author{Thomas Barois$^*$, M\'elanie Labiausse, Jean-Fran\c cois~Boudet\footnote{Our thoughts are with Jean-Fran\c{c}ois Boudet who passed away while this manuscript was under review.}, Juho S. Lintuvuori, and Hamid Kellay \\
\email[Corresponding author: ]{thomas.barois@u-bordeaux.fr}\\
\normalsize{Univ. Bordeaux, CNRS, LOMA, UMR 5798, F-33400 Talence, France,}\\}
\begin{abstract}
We experimentally and numerically demonstrate the self-assembly of granular rotors into chain-like structures. The rotors are non-convex rigid bodies with rotational activity. We first investigate the interaction between two rotors and we establish an attractive interaction controlled by the ratio of the rotational activity to the noise and also dependent of the rotors geometry. We then show that the cohesive state between spinner pairs leads to robust chains that can be either initially set or spontaneously obtained by self-assembly.  \end{abstract}
\maketitle
\section{Introduction}

Active granular matter provides a physical framework for the emergence of collective behaviours with ensembles of motile particles\cite{vicsek1995novel,Narayan2007,Kudrolli2008,Kumar2014,peruani2016active}. In active granular media, self-organized ordered states\cite{Deseigne2010,li2013asymmetric,deblais2018boundaries,kumar2019trapping,riedel2024designing} can emerge under the effect of aligning interactions caused by steric considerations and collision dynamics.     

Active granular particles are typically self-propelled particles, driven by vibrations, and with a rectifying mechanism that arises from the particles asymmetry\cite{ioi1999mobile,yamada2003coherent,dorbolo2005dynamics,Giomi2013,xu2017self}. The motility of active particles can be purely translational in case of polar particles or a combination of translational and rotational activity for chiral particles. Chiral active particles can exhibit circular trajectories\cite{van2008dynamics,radtke2012directed,kummel2013circular,ai2016ratchet,liebchen2017collective,lei2019nonequilibrium}, when translation and rotation are both present, or a steady spinning\cite{kokot2017active,farhadi2018dynamics,workamp2018symmetry,shen2019hydrodynamic,yang2020robust,liu2020oscillating,shen2023collective}, if there is no translational activity. Chiral active particles are of notable interest for the variety of behaviours they generate as a bulk ensemble\cite{tsai2005chiral,van2016spatiotemporal} and also for presenting phase separation and sorting mechanisms\cite{mijalkov2013sorting,nguyen2014emergent,chen2015sorting,yeo2015collective,liebchen2016pattern,scholz2018rotating,levis2019activity,barois2020sorting}. 

Here we explore a new strategy of self-assembly for granular particles based on chirality.
Although granular particles interact by repulsive interactions, we show that with dumbbells geometry, an effective attractive force can emerge from chiral activity. We  quantify the interaction potential between two isolated rotating dumbbells and we show that the  attractive interaction can collectivelly lead to self-organized stable chains, a type of collective state that has not been reported so far in active granular media. 

The self-organization of elements in chains is a universal building block found at all scales, with the example of polymers at the molecular scale, microtubules\cite{martin1990template} and fibers\cite{lenz2017geometrical}  in biophysical systems, fungal networks\cite{alim2013random}, or trails and living bridges of ants\cite{reid2015army,hu2016entangled}. With our strategy to form chains with chiral granular elements, we open a new toolbox that can be combined with the already existing strategies of self-assembly in swarm robotics\cite{yan2012linking,rubenstein2014programmable,fischer2018machine,li2019particle,savoie2019robot,li2021programming,molodtsova2025micellization,wang2025hyperuniform}
 and multimodal locomotion\cite{dorbolo2009bouncing,kubo2015mode,won2019demand,shelke2019transition,barois2024controlled}

\section{Clustering of granular rotors}
\begin{figure}[h!]
    \centering
    \includegraphics[width=0.5\linewidth]{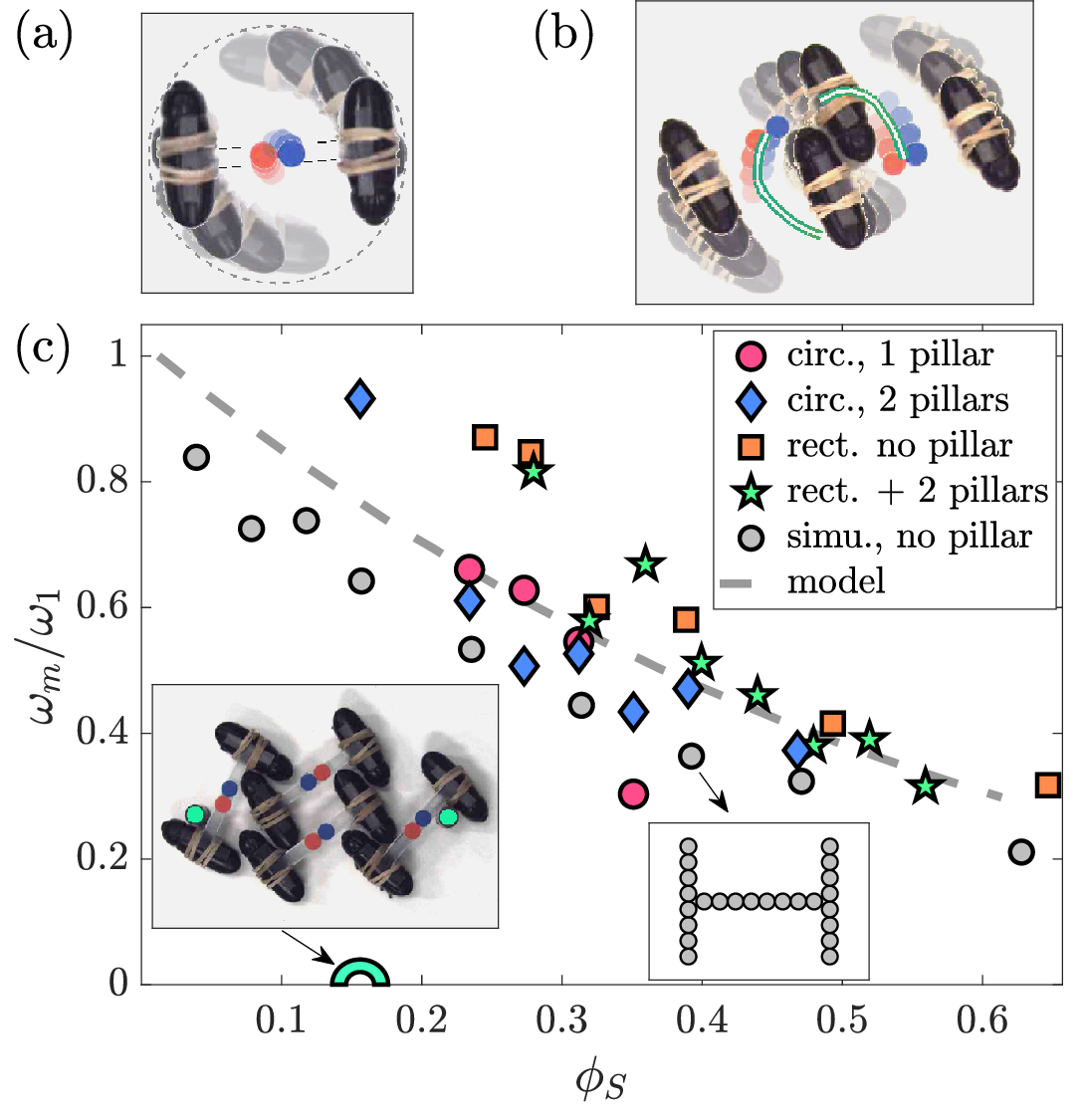}
    \caption{a) Rotor in a free-spinning regime.  b) Two rotors paired in a steady orbiting structure. c) Angular frequency of a set of rotors as a function of their surface density $\phi_S$ for various confinements with $\omega_1$ the angular frequency of solitary rotors. With two pillars (green dots in the inserted image), it is possible to have stable chains, as shown in the inserted view with 4 steady rotors (bridge symbol, $\omega_m = 0$). circ. and rect. refer to circular and rectangular enclosure respectively. The simulation is performed with a H-like rotor made of unit spheres   (see appendix \ref{app:simu}).}
    \label{fig:omega_phi}
\end{figure}

The rotor we use in this work is presented in figure \ref{fig:omega_phi} a. It is obtained by pairing two self-propelled robots with an opposite orientation, linked by a rigid rod. In figure \ref{fig:omega_phi}, the robots are all paired such that they perform a clockwise rotation. When isolated as in  figure \ref{fig:omega_phi}a (movie M1), the rotors are in a steady rotation with a diffusing motion of their center of mass and of their phase. When two rotors of the same polarity are brought together as in figure  \ref{fig:omega_phi} b (movie M2), they can form a binary orbiting structure, with an angular rotation  that is typically half of the free angular rotation of the isolated rotors. This binary state is only observed for pairs of rotors of the same rotation sign. Contrarily to the unpaired robots\cite{deblais2018boundaries}, the orientation dynamics of the rotors with two paired robots is in the overdamped regime.

In figure \ref{fig:omega_phi} c, we represent the mean angular frequency $\omega_m$ of an assembly of $N$ rotors placed in a box of surface $S$. $\omega_1$ is the angular frequency of free rotors and it is of the order of 7.0 rad/s in experiments. The surface fraction $\phi_S$ is the ratio of the rotors area $N\times S_1$ divided by the total surface area $S$. $S_1$ is the bounding circle of a free rotor as presented in figure \ref{fig:omega_phi}a. This definition of the particle area makes possible the regime $\phi_S > 1$. We eventually place static pillars in the box: the pillars tend to slow down the rotors spinning velocity but the main point behind the presence of pillars is to enable the formation of long-lasting chains of rotors by providing anchoring points. At low packing fractions, the rotors mostly diffuse and rarely form binary structures. At larger surface fractions, structures composed of two or more rotors tend to form  which results in a slowing down of their angular rotation (movie M3). The solid line in the figure is a statistical model that counts the average number of clusters of size $N_c$ for randomly distributed particles assuming an angular velocity depending on the cluster size $\omega_{N_c} \propto 1/N_c$ (see Appendix \ref{app:model_omega_phi} for more details). The simulation is performed with H-like rotors (see Appendix \ref{app:chain_stat}). Although there is no pillar for the simulation presented in figure \ref{fig:omega_phi} c, $\omega_m/\omega_1$ is typically smaller than in the experiments, which we attribute to the sharper shape of the H-like rotors compared to the experimental rotors.

 The first message in figure \ref{fig:omega_phi} c is the decrease of the rotors spinning velocity with the increase of density. This is a crowding effect that comes from the increasing number of rotor clusters forming with increasing surface density. The second and more important message in figure \ref{fig:omega_phi} c is existence of chain-like structures, represented by the data point at $\phi_S = 0.15$ ($N=4$) and $\omega_m=0$. This regime of arrest is only observed with 2 pillars to which the rotors can attach to form  persisting chains (movie M4). Without pillars, there are no steady chains observed, apart from the orbiting chains with $N_{ch}=2$ in the rotating frame.
 For the two data points at $\phi_S=0.15$, the only difference is the initial configuration (diamond symbol: no chain at the initial state, bridge symbol: the 4 rotors are manually positioned as a chain). This chain is a mechanically stable configuration and it was maintained during the duration of the acquisition (3 minutes). The distance between the pillars requires a minimum number of 4 rotors to form a connecting chain. For $N=4$ (diamond symbol), we have not observed the spontaneous formation of a chain during the duration of the experiment. For a larger number of rotors, we observe multiple events of spontaneous chain formation. The bridge between the pillars can last for a certain time before it dissolves by the disturbance of the surrounding rotors, notably if $N$ is significantly larger than the number of rotors in the chain. With $N=6$, we observed a spontaneous chain formation with 4 rotors and the two remaining rotors were not able to destroy the chain that was maintained until the end of the experiment (see movie M5 for a spontaneous chain formation with $N=9$ rotors).

\section{Pair interaction of model rotors}
\subsection{Binding of spherical dumbbells}
\begin{figure}[h!]
    \centering
    \includegraphics[width=0.7\linewidth]{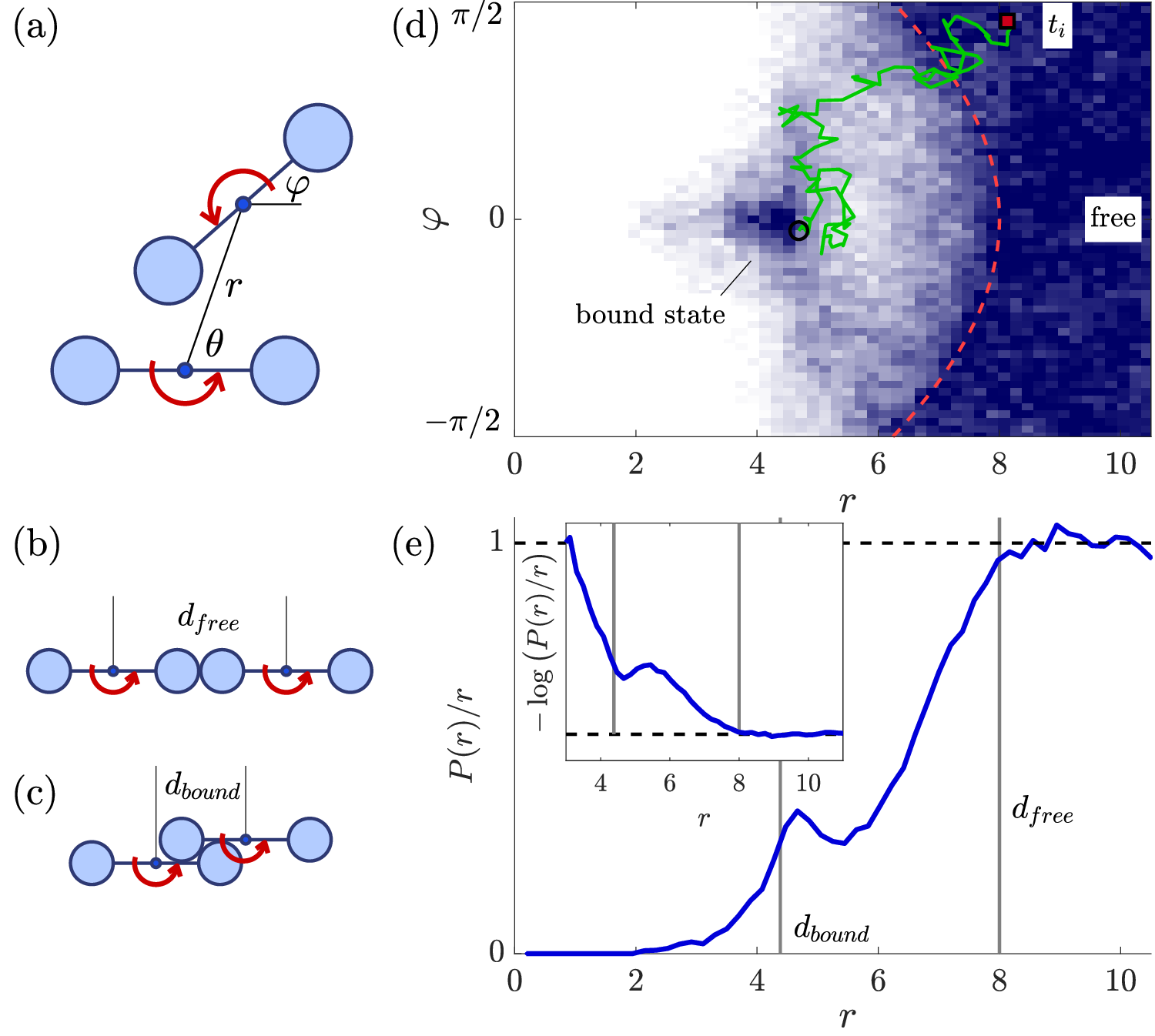}
    \caption{a) Representation of two dumbbells at distance $r$, angular position $\theta$ and orientation difference $\varphi$. The arrows indicate the self-imposed torque on each dumbbells. The radius of the sphere is $R=1$ and the distance between the sphere is $L=6$ b) Configuration of farthest contact $r=d_{free}$, $\varphi =0$. c) Configuration of the bound state with $r = d_{bound}$. d) Probability of presence of two interacting torque-forced dumbbells as a function of their distance $r$ and orientation difference $\varphi$. e) Probability of presence $P(r)$ of two dumbbells at a distance $r$ ($\varphi$-average of figure d).  The inserted plot is $-\log \big(P(r)/r\big)$.}
    \label{fig:model_dumbbell}
\end{figure}

To address the binding mechanism for pairs of rotors, we present in figure \ref{fig:model_dumbbell} a model of torque-driven dumbbells with  repulsive contact interaction. The dumbbells are composed of two spheres of radius 1 linked by a line segment. The distance between the spheres is $L=6$. The configuration space of two interacting dumbbells can be described by 3 independent variables represented in figure \ref{fig:model_dumbbell} a : the center-to-center distance $r$,  the cylindrical coordinates angle $\theta$ and the orientation difference $\varphi$ between the two dumbbells. 
For each set of $\{r, \theta, \varphi \}$, the configuration is either allowed or forbidden if the two dumbbells overlap. 
By a molecular dynamics approach, we sample the accessible configurations of two dumbbells under Brownian diffusion and rotational activity (movie M6).   
In figure \ref{fig:model_dumbbell} d, we present the probability of presence of two interacting dumbbells in the space $\{r, \varphi\}$. This contour plot is normalized by a function $f(r)\sim r$ so that non-interacting particles are represented with the same color depth (see appendix \ref{app:eff_pot}). 
Outside the region delimited by the curved dashed line (free region), the distance $r$ between the two dumbbells is large enough so that for any $\theta$, the dumbbells do not overlap and all the space $\{r, \varphi\}$ is equally accessible.  In figure \ref{fig:model_dumbbell}b, the point $r=d_{free}=8$, $\varphi = 0$ marks the limit of the free region. Inside the dashed line, not all the configurations are equally accessible because of the constraint on the variable $\theta$ and also because the dumbbells activity favors some binding configuration like the one presented in figure \ref{fig:model_dumbbell}c. In figure \ref{fig:model_dumbbell}d, the point at $\{r=d_{bound}  = \sqrt{(L-\sqrt{3}R)^2 + R^2}\approx 4.38$, $\varphi=0\}$ is an attractor and we show a sample trajectory of the dumbbells dynamics starting at time $t_i$ at the position $r=8$, $\varphi = \pi/2$ and ending at time $t_f$ at $d_{bound}$. The bound state is localised at the vicinity of $\varphi = 0$ which means that the cohesion is strongly anisotropic and favors a preferred orientation for the spinners compatible with chain-like structures. 

In figure \ref{fig:model_dumbbell}e, we represent the probability of presence $P(r)$ of the two dumbbells as a function of the distance $r$. $P(r)$ is normalized by $r$ to account for the geometry of cylindrical coordinates $\{r, \theta\}$.
$P(r)/r$ is obtained by averaging the probability map in figure \ref{fig:model_dumbbell}d over $\varphi$. The plot shows the distance of farthest contact $d_{free}$ after which the probability is constant and local maxima at $d_{bound}$ that shows that the configuration at $d_{bound}$ is an attractor. The point of closest approach is $r = 2$ below which the probability of presence is strictly zero. The fact that the probability locally decreases when $r$ increases in the range 4.5 to 5.5 indicates an effective attraction induced by the self-rotation of the dumbbells. In appendix \ref{app:pot_brownian}, we provide the equivalent for figure \ref{fig:model_dumbbell}d and e in the Brownian limit (ie no forced rotation) and we establish theoretical solutions for the probability of presence for $\varphi = 0$ and $\varphi = \pi/2$ as a function of $r$. In the inserted plot of figure \ref{fig:model_dumbbell}e, we represent $-\log (P(r)/r)$ with the idea that the probability distribution $P(r)$ could be used to determine an effective interaction potential $U_{{eff}} \propto -\log (P(r)/r)$ between two rotors. 

\subsection{Effective interaction potential with spherocylinder dumbbells}
\begin{figure}[h!]
    \centering
    \includegraphics[width=0.7\linewidth]{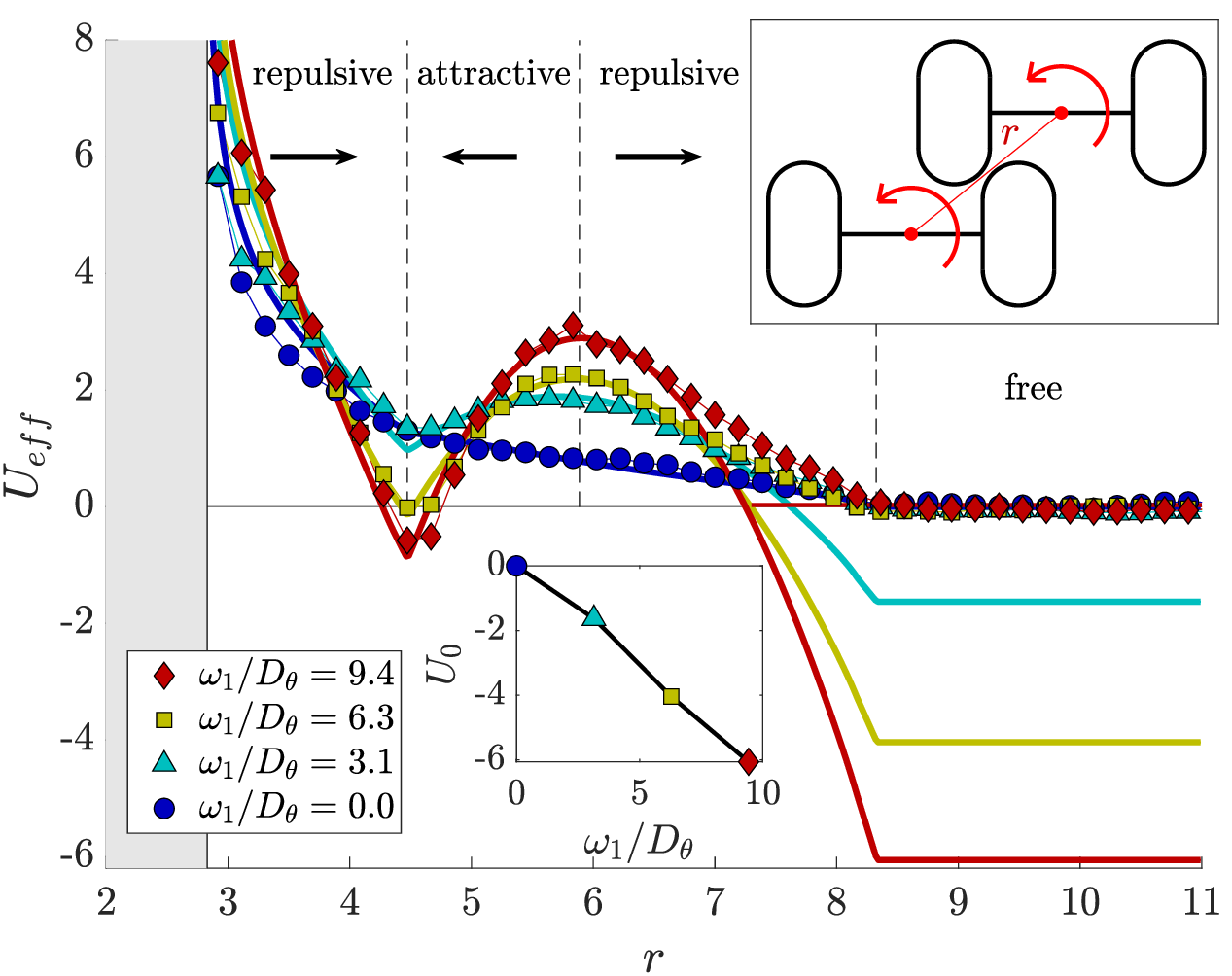}
    \caption{Effective interaction potential $U_{{eff}} = -\log P(r)$ of two interacting rotors with the geometry represented in the inserted top-right panel as a function of the rotational activity $\omega_1/D_\theta$, with $\omega_1$ the angular frequency of the free rotors and $D_\theta$ the rotational diffusion. The symbols are the numerical results of a Langevin dynamics with rotational activity. The solid lines are the mechanical model $U_{{Brownian}}(r) + U_{{Active}}(r)$. The vertical dashed lines are obtained from the rotor geometry. $U_0$ is an added parameter of the model potential $U_{eff}$ that corresponds the value of the plateau at $r>8.3$. }
    \label{fig:pot_energy}
\end{figure}

In figure \ref{fig:pot_energy}, we explore the idea of an effective interaction potential between rotating dumbbells. In order to approach the geometry of the experiments, we use spherocylinder elements, ie a square body of size 2, closed by two half circles of radius 1. 
From the probability of presence $P(r)$ at a distance $r$, we infer an effective potential $U_{{eff}} \propto -\log(P(r)/r)$ that we compute for 4 conditions of rotational activities $\omega_1/D_\theta$ with $\omega_1$ the free frequency of the rotors and $D_\theta$ the angular diffusion coefficient. $D_\theta$ is measured from the orientational diffusion of an isolated rotor without activity. The plot at $\omega_1/D_\theta = 0$ is the Brownian limit without rotational activity. With the rotational activity increase (triangle, square and diamond symbols), the interaction potential shows a more pronounced attraction in the range $r=4.5$ to $r=6$. 
%\begin{equation}
%    U\left(r,\frac{\omega_0}{D_\theta}\right) = U_{Brownian}(r) +   U_{Active}(r) + U_0\label{eq:model}
%\end{equation}

The physical mechanism at the origin of the attraction  can be understood with the two represented rotors in the inserted box in figure \ref{fig:pot_energy}. The distance between the two rotors is $r=5.4$ and under the effect of the rotation, the rotors will come closer in a bound configuration similar to the one of figure \ref{fig:model_dumbbell}c with $d_{bound} \approx 4.5$. To capture this physical mechanism, we propose a model for the interacting potiential $U_{{eff}} = U_{{Brownian}}(r) + U_{{Active}}(r)$
in which $U_{{Brownian}}(r)$ is the potential in the absence of rotational activity, $U_{{Active}}(r)$ is the contribution from the active rotation. The Brownian potential is the blue solid line in figure \ref{fig:pot_energy}. It is numerically computed assuming an equal distribution of all the two-dumbbell accessible configurations. We compute the probability of presence as a function of $r$ by averaging the distribution over $\theta$ and $\varphi$ and we infer the potential $U_{Brownian}$ by the Boltzmann relation.

%To compute the part of the interaction potential due to the rotational activity $U_{Active}(r)$, we propose a mechanical model that considers the rotors as frictionless elements. We consider the contact configuration for which the two rotors have the same orientation ($\varphi=0$) and we estimate the magnitude of the force $F_u$ that would be needed to avoid the sliding of the two particles under the effect of the rotational activity.
To estimate $U_{{Active}}(r)$, we use a mechanical model that evaluates the magnitude of an additional force $F_u(r)$ that would be required to maintain two dumbbells in contact if they were in solid body rotation at a distance $r$ with the same orientation $\varphi=0$ (see appendix \ref{app:model_interaction}). The force $F_u$ is  parallel to the sliding direction of two dumbbells in contact.  
 Using the geometrical properties we have:
\begin{eqnarray}
    F_u(d_{min} < r < d_{bound}) &=& -\frac{4\Gamma R}{r^2 + L^2}\\
    F_u(d_{bound} < r < d_{free}) &=& \frac{2\Gamma}{r^2 + L^2}\frac{L^2 - r^2}{4R}
\end{eqnarray}
with $d_{min} = \sqrt{2^2 + 2^2}\approx 2.83$ the minimal distance of approach, $d_{bound}= \sqrt{4^2 + 2^2}\approx 4.47$ the binding distance and $d_{free}\approx 8.32$ the farthest distance of contact. $\Gamma$ is the torque applied to the particles, $L$ is the length of the rotor and $R$ is the radius of the spherocylinders. 
For $d_{min} < r < d_{bound}$, the interaction is repulsive. For $d_{bound} < r < d_{free}$, the force changes sign at $r=L$ and this term adds an attractive interaction for $r<L$ and a repulsive interaction for $r>L$.  For $r > d_{free}$, the model is ill-defined because the dumbbells are not in contact.
From  $F_u$, we can integrate  the potential using the relation $U_{{Active}}(r) = -A\int_r^\infty F_u(r)dr + U_0$. The dimensionless parameter $A$ sets the relative weight of the two potentials $U_{{Brownian}}$ and $U_{{Active}}$. This parameter is set to a unique value. $U_0$ is an additional integration constant set to different values depending on $\omega_1/D_\theta$, as shown in the inserted plot in figure \ref{fig:pot_energy}. Our model is in remarkable agreement with the simulation results in the range $0<r<7$. For $r>7$, our model fails to predict the free regime but this is not surprising because our model assumes contact between the two dumbbells. We can however note that $U_0 \sim -\omega_1/D_\theta$ and we interpret this dependence as a consequence of the periodic interaction between free dumbbells (see appendix \ref{app:interpret}). 
%To rationalize the physical meaning of this offset term, we can argue that we have used an at-equilibrium approach to infer a potential from the distance probability distribution which is \textit{a priori} non valid for active particles. Actually, the rotors are out of equilibrium and we can interpret the term $U_{offset}$ as an additional source of thermal noise coming from the rotational activity: if we want to establish a characteristic fluctuating velocity of the rotors, it must be a combination of the Brownian velocity of the particles and the characteristic velocity difference between two interacting rotors.

\section{Chain formation}
\begin{figure}[h!]
    \centering
    \includegraphics[width=0.5\linewidth]{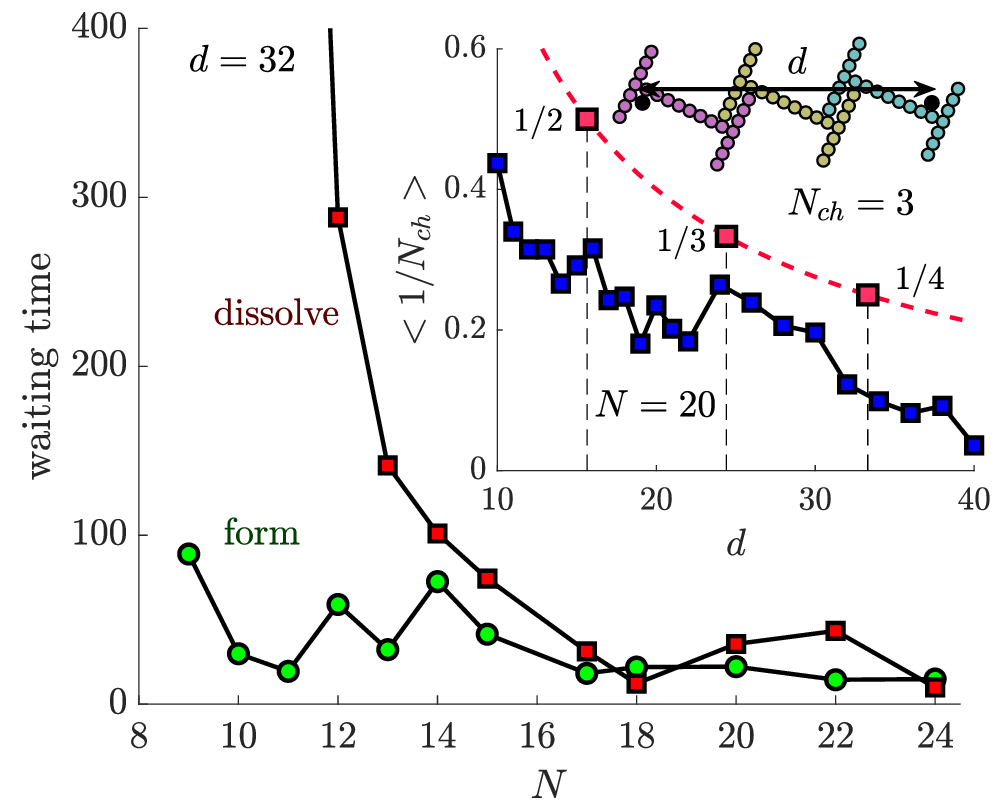}
    \caption{Simulation results for the waiting time to form and dissolve a chain between two pillars at a distance $d$ with H-shaped rotors. 
    The time is expressed in rotation time units and the total simulation time is 347. For $N = 9$, 10 and 11, the dissolution time is larger than 1000 periods.  The surface fraction for 1 rotor is 0.018 ($N=8$ is $\phi = 0.14$ an  $N=24$ is $\phi = 0.43$). The inserted plot shows the analogue of a conductivity by the time average $<1/N_{ch}(t)>$ between the two pillars with $N_{ch}(t)$ the number of rotors connecting the 2 pillars at time $t$. If there is no chain at  time of the frame, $1/N_{ch}$ is 0. The simulation is performed with $N=20$ ($\phi_S = 0.359$). The dashed line is an upper bound for $<1/N_{ch}>$ assuming a linear chain.}
    \label{fig:simu_chains}
\end{figure}

An important result of this paper is the spontaneous formation of chains that we observe both in experiments and simulations. To explore the chains dynamics, we simulate rotors composed of unit spheres placed on a H-like backbone. The size of the rotors is $11\times 9$ (see appendix \ref{app:chain_stat}). The rotors are in a circular domain of diameter 100, with two static pillars at a distance $d$. In figure \ref{fig:simu_chains}, we present in the main plot the waiting time for chain formation and dissolution as a function of the number of rotors for a distance $d=32$ between the pillars. The waiting time is expressed in rotation periods of a free rotor. The first occurrence of chain formation is for $N = 8$ ($\phi = 0.14$) after a waiting time of 538 rotation periods. This data point at $N=8$ is not represented in figure \ref{fig:simu_chains} because no breaking of the chain was observed. In the regime between $N=8$ and $N=13$ ($\phi = 0.23$), the formation time is significantly smaller than the dissolution time which means the chain persists. For $N>13$, the chain formation time remains small but the dissolution time also shortens which is due to the large number of rotors disturbing the chains.

The inserted plot in figure \ref{fig:simu_chains} analyses the connectivity between two pillars separated by a distance $d$ for 20 rotors during a simulation. We evaluate $<1/N_{ch}(t)>$ for the shortest path linking the two pillars in which the number of rotors in the chain $N_{ch}(t)$. $<1/N_{ch}(t)>$ quantifies how reliable is the connection between the two pillars during time: a larger value of $<1/N_{ch}(t)>$ means that a short and time-persistent chain is present. We average $1/N_{ch}(t)$ for each time frame of the simulation and if there is no chain,  $1/N_{ch}(t)$ is set to zero. The dashed line is an extrapolation of the regime  $<1/N_{ch}> = 1/i$ with $i$ a positive integer. We observe local maxima at the distances associated to 1/2 and 1/3 which is consistent with pillar distances that can accommodate an integer number of rotors, with for example $N_{ch}=3$ and $d=24$ as shown in the inserted view. 

%We used the numerical simulation of active rotors to test the behaviour of an equal number of clockwise and counterclockwise particles in a circular container. We found that the mixture separates into small dynamical structures of typically 5 to 6 particles with the same polarity. 

To have stable chains with more than two rotors, we used, in experiments and simulations, static pillars to anchor the free end of the chain. It is however possible to have stable chains  by connecting a manually-made chain to itself to form a stable loop (movie M7). By exploring the shape of the rotors, we were also able to produce steady chains for more than $N=2$ with X-shaped particles. In movie M8 we show that small chains of few links can be obtained by self-assembly and in movie M9 we show that there is no restriction to the stability of long chains without pillars by manually forming a chain of $N=20$.

\section{Conclusion}
In this paper, we showed that granular rotors with contact interaction can form bound states in the form of orbiting pairs of particles or linear chains with many rotors. For the 2-rotor situation, we established an effective interaction potential constructed using a mechanical model that depends on the geometry and the activity of the rotors. For multiple rotors, the obtained chains are collective structures that can emerge from self-assembly. 
To show the universality of our observations, we have explored a class of particles geometry derived from spherical dumbbells that gives the possibility to vary the level of non-convexity from moderate (spherical dumbbells) to large (spherocylinders, H-like rotors). 
This work shows how simple granular entities can be used to design self-assembling chain-like structures which are essential elements in building processes.

\section*{Acknowledgments}
The authors thank Thomas Gu\'erin for scientific discussions related to this work. We acknowledge the support of the French Agence
Nationale de la Recherche (ANR), under Grant No. ANR-22-CE06-0007-02.

\appendix
\renewcommand{\thefigure}{A\arabic{figure}}
\setcounter{figure}{0}
%\counterwithin{figure}{section}
\section{Experimental setup}
\subsection{Angular velocity of the rotors}
\begin{figure}[h!]
    \centering
    \includegraphics[width=\linewidth]{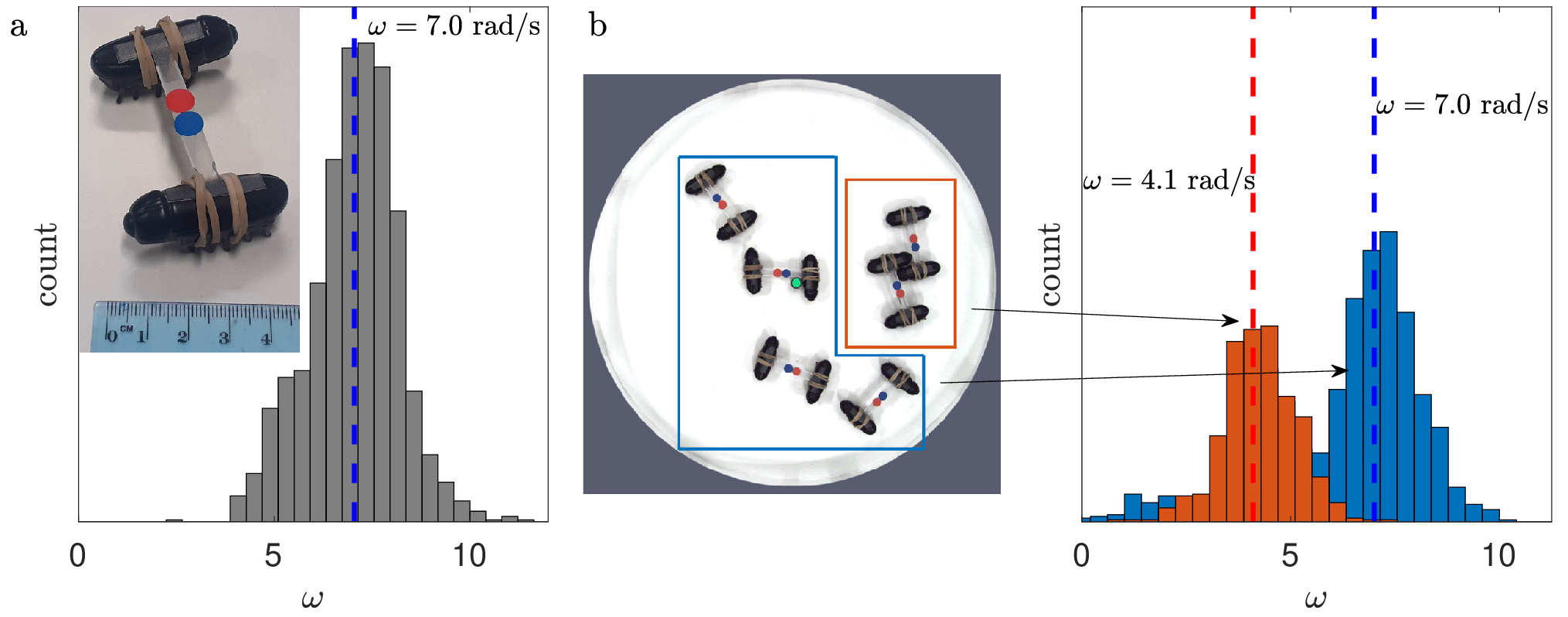}   
    \caption{a) Angular velocity distribution of $N=6$ rotors without contact interactions. b) Angular velocity distribution for two sets of rotors: isolated rotors (blue) and paired rotors (orange). }
    \label{fig:rotors_omega_distr}
\end{figure}

The experimental rotors are obtained by connecting two self-propelled robots with a bridging structure on which two stickers are glued. In the configuration of figure \ref{fig:rotors_omega_distr} a, the rotor spins in the clockwise direction. The robots could also be connected to have a counterclockwise rotor. The length of the robots is 4.5 cm and the size of the rotor is of the order of 7 cm. A red and a blue stickers are glued on the linking segment of the rotors to track their solid body motion. 
%\begin{figure}[h!]
%    \centering
%    \includegraphics[width=0.5\linewidth]{photo_bugs.eps}   
%    \caption{Image of an individual active robot and of a rotor composed of a pair of robot linked by a plastic layer with two stickers used for tracking analysis.}
%    \label{fig:rotors_exp}
%\end{figure}

%\subsection{Angular velocity}
The free-rotation frequency is self-chosen by the rotors and varies from one element to another and also varies with the discharging of the robots battery. In figure \ref{fig:rotors_omega_distr} a, we show the angular velocity distribution for 6 non-interacting rotors. 

The angular velocity distribution for two bound rotors is presented in figure \ref{fig:rotors_omega_distr} b with the distribution for the two paired rotors in orange with maximal angular velocity probability at $\omega = 4.1$ rad/s and in blue for 4 other rotors with a maximal angular velocity probability at 7.0 rad/s. 
%\begin{figure}[h!]
%    \centering
%    \includegraphics[width=0.55\linewidth]{distr_omega_N6_paires.png}
%    \hfill
%    \includegraphics[width=0.4\linewidth]{image_paire_from_166_1960_cadre.png}
%    \caption{Angular velocity distribution for two sets of rotors: isolated rotors (blue) and paired rotors (orange). }
%    \label{fig:rotors_distr_omega_pair}
%\end{figure}

\subsection{Clusters dynamics}
\begin{figure}[h!]
    \centering
\includegraphics[width=0.45\linewidth]{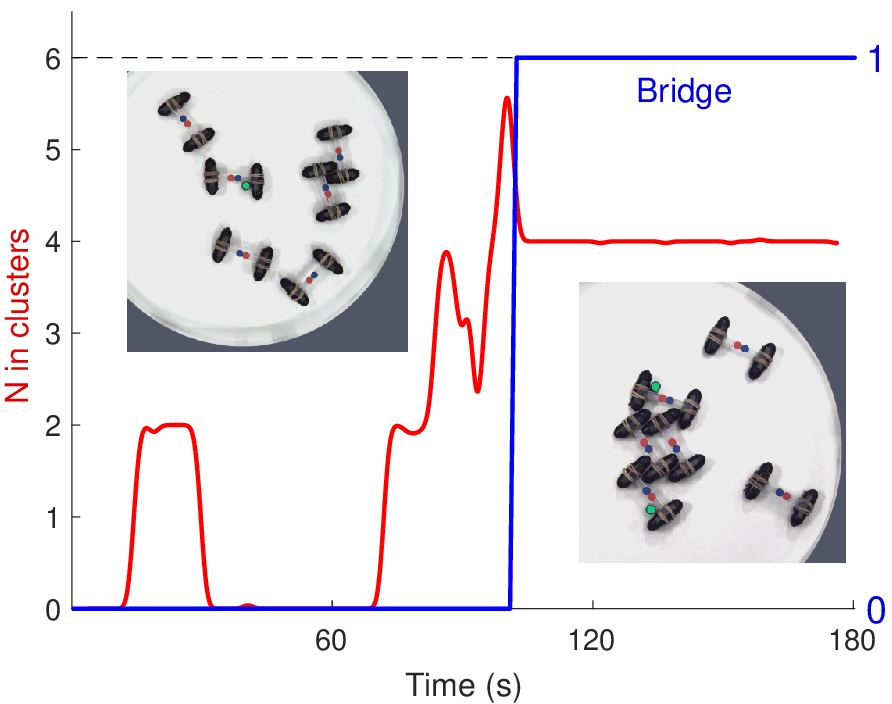}
    \caption{Number of rotors belonging to a cluster as a function of time (red solid line) in an experiment with $N=6$ rotors. The blue solid line is a binary function that outputs the presence of a linking bridge between the two static pillars by determining if a cluster is in simultaneous contact with the two pillars.}
    \label{fig:cluster_dynamics}
\end{figure}

A typical experiment consists in placing $N$ rotors in an enclosure and extracting the trajectory of the particles by tracking of the stickers on their body. Because we have access to the position and orientation of the rotors, we can determine which set of rotors are in a collective solid body rotation, thus defining a cluster of aggregated particles. Figure \ref{fig:cluster_dynamics} shows the number of rotors belonging to a cluster of any size as a function of time for a total number of 6 interacting particles. At $t= 25$ s there is one cluster with 2 particles (the solid body rotation of this cluster is of the order of 4.1 rad/s, see figure \ref{fig:rotors_omega_distr} b) . After $t=100$ s a larger cluster of 4 particles and the solid body rotation of this cluster is zero because the cluster forms a steady chain between two static pillars in the arena. 

\section{Model for the angular velocity as a function of surface density of rotors}\label{app:model_omega_phi}
\begin{figure}[h!]
    \centering
    \includegraphics[width=0.5\linewidth]{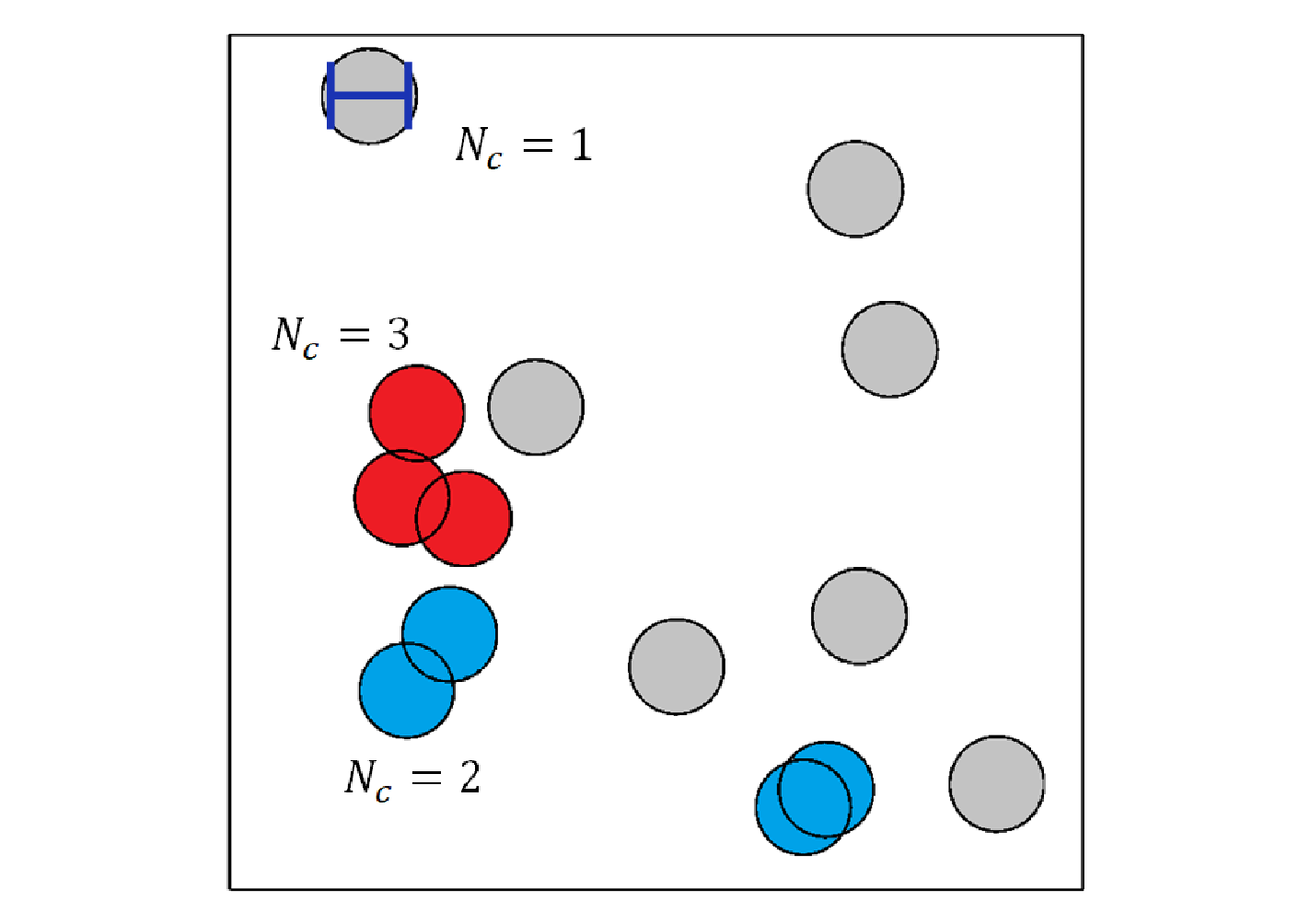}
    \caption{Representation of the statistical model for the mean angular velocity. 14 particles are randomly placed in a box of size $L_b=16$. The cluster size distribution is obtained by associating a disc of radius 1 to each particles.}
    \label{fig:disc_model}
\end{figure}

In figure \ref{fig:omega_phi}, we use a statistical model to predict the evolution of the normalized angular velocity of the rotors. In this model we consider a box of size $L_b=16$ and $N$ point-like particles randomly placed in the box. For each particle, we associate a circle of radius 1 and we compute the clusters by associating together contacting circles as shown in figure 
\ref{fig:disc_model}.

\begin{figure}[h!]
    \centering
    \includegraphics[width=0.5\linewidth]{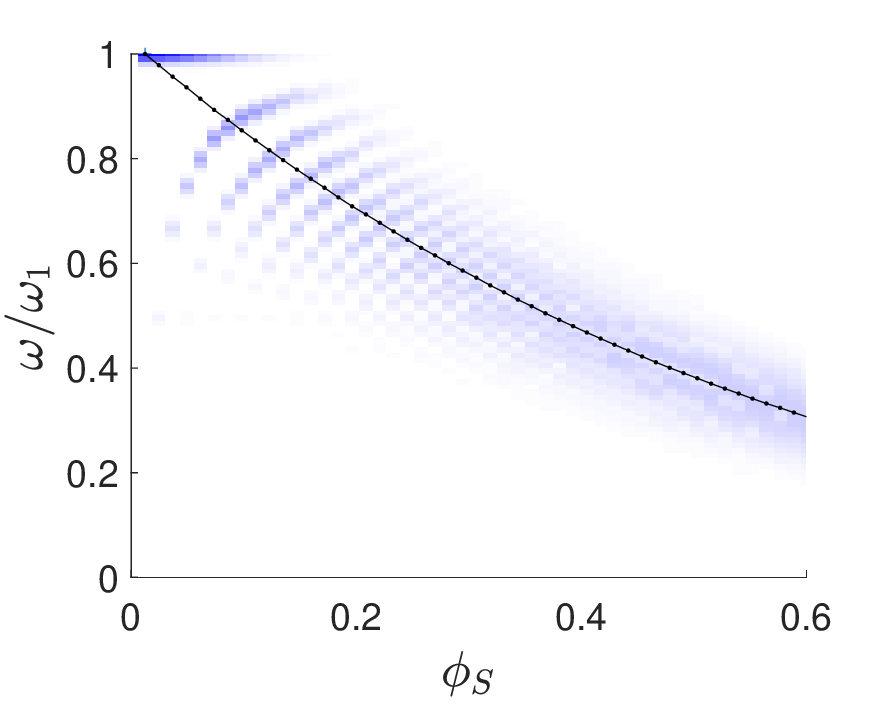}
    \caption{Normalized angular velocity as a function of the surface fraction obtained from the Monte-Carlo simulation of a model of random distribution of particles assuming a relation $\omega_{N_c} \propto 1/N_c$ for particles in clusters of size $N_c$. The blue fringes are the contribution of cluster of 1, 2, 3, etc with angular velocities 1, 1/2, 1/3 respectively.}
    \label{fig:omega_phi_simu}
\end{figure}

The model assumes that the angular velocity of a cluster of size $N_c$ is $\omega_{N_c} \propto 1/N_c$. To obtain the mean angular velocity of $N$ particles in a box of size $L_b$, we sum the angular velocity from each particles using the size of the cluster they belong to and we repeat the particles configuration by a Monte Carlo method with 50000 realisations for each $N$. The result of the simulation for the model is presented in figure \ref{fig:omega_phi_simu}. 

\section{Numerical simulation}\label{app:simu}
\subsection{Simulation method}
\begin{figure}[h!]
    \centering
    \includegraphics[width=0.5\linewidth]{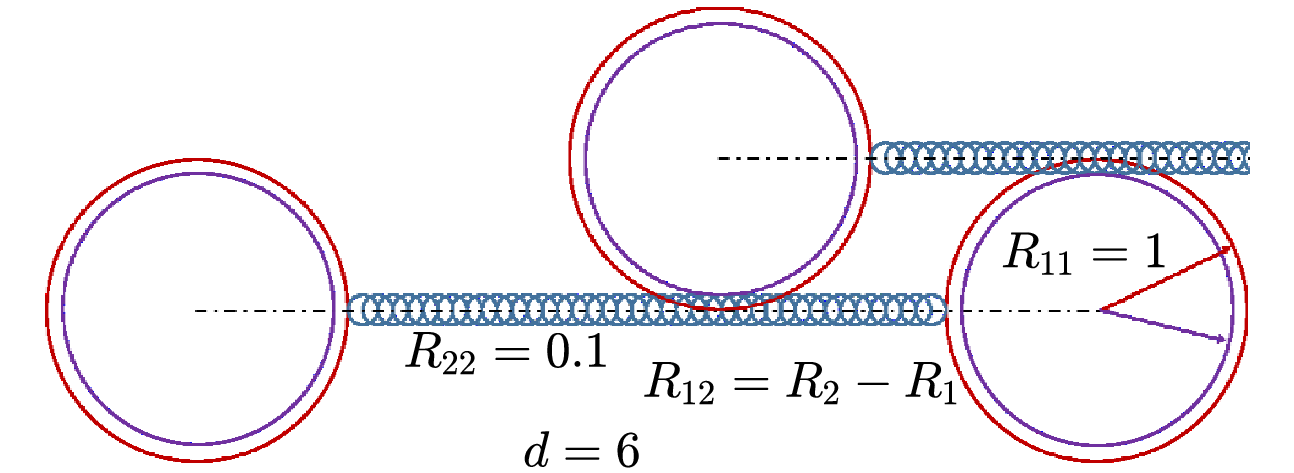}
\caption{Spherical dimers with 2 spheres of radius 1 per dimers and 71 smaller spheres of radius 0.1 used to approximate the line segment linking the two spheres.}
    \label{fig:dimers_sphere}
\end{figure}

We use the molecular dynamics software LAMMPS to simulate the dynamics of the rotors. Each  rotor is described by a set of rigid spheres assembled in a rigid body structure with short-range repulsive interaction. Figure \ref{fig:dimers_sphere} shows a dumbbell made of two spheres, labeled 1, of unit radius separated by a distance 6 and a set of spheres, labelled 2, of radius 0.1 to form the rigid link between the spheres. The interaction between particles is a truncated Lennard-Jones potential with its repulsive part only. We use a characteristic distance 1 for the repulsion between the sphere of radius 1, 0.1 for the spheres of radius 0.1 and 0.9 between the sphere of type 1 and 2 to mimic a linking segment of vanishing width (see figure). 

\begin{figure}[h!]
    \centering
    \includegraphics[width=0.5\linewidth]{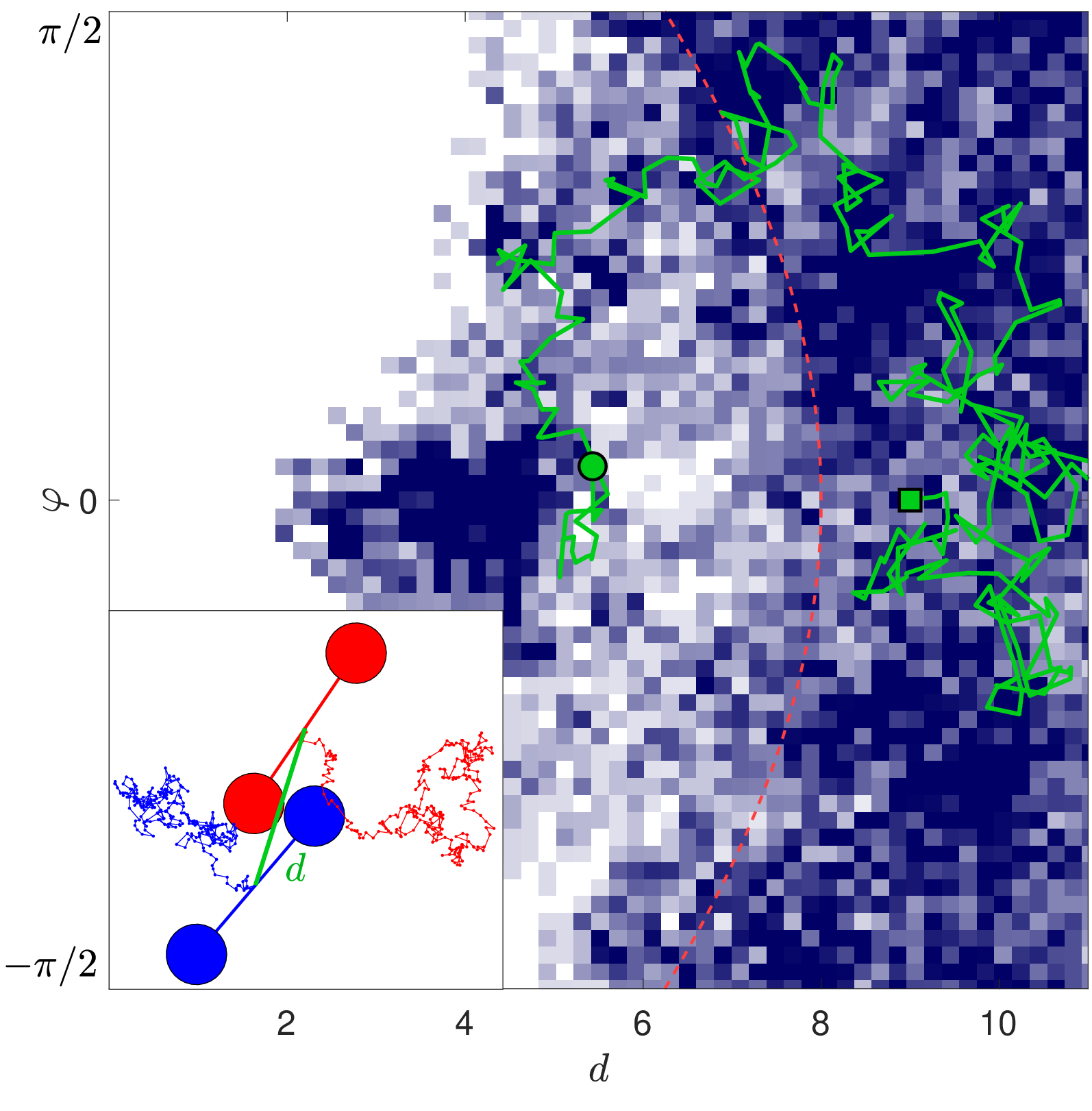}
    \caption{Probability map of the two-dimer configuration with $d$ the distance between the dimers and $\varphi$ the orientation difference between the two dimers orientation. The green line is a sample trajectory used for the binning process.  Inserted plot: Trajectories of two rotors with Brownian noise (the tracks follow the center of each dimer and $d(t)$ is the distance between the dimers centers.}
    \label{fig:brownian_dimers_map}
\end{figure}

The simulation solves a Langevin equation for a set of spherical particles that forms a rigid body object. For each spherical particle, the Langevin equation has the generic form:
$$
m \frac{d}{dt}v + \gamma v = F_a + F_i 
$$
in which $m/\gamma$ is a relaxation timescale, $F_a$ is an active force and $F_i$ is the interaction force with the other particles. For the dumbbell geometry, the applied active forces are equivalent to an external torque applied to rigid body.
We restrict our numerical study to the dynamics of the rotors in the overdamped limit ($m/\gamma=
0.01$ time units). Each rotor is submitted to a Brownian-like noise and an external torque (or equivalently pair of active forces with opposite direction). The balance of viscous friction and applied torque sets the typical rotation frequency of the rotors. Figure \ref{fig:brownian_dimers_map} represents the distribution for $2\times 10^5$ rotation period  and the density contour in constructed by binning of the trajectories in the space $\{d,\varphi\}$, in which $d$ is the distance between the dumbbell centers and $\varphi$ is the angle difference between the two dumbbells ($\varphi$ is close to zero in figure \ref{fig:brownian_dimers_map}). In the inserted plot of figure \ref{fig:brownian_dimers_map}, we represent the simulated trajectory of two dumbbells in a periodic box of size 22 space units. The dumbbells rotate counterclockwise and their period of rotation is of the order of 100 time units. 

\subsection{Box size}
\begin{figure}
    \centering
    \includegraphics[width=0.5\linewidth]{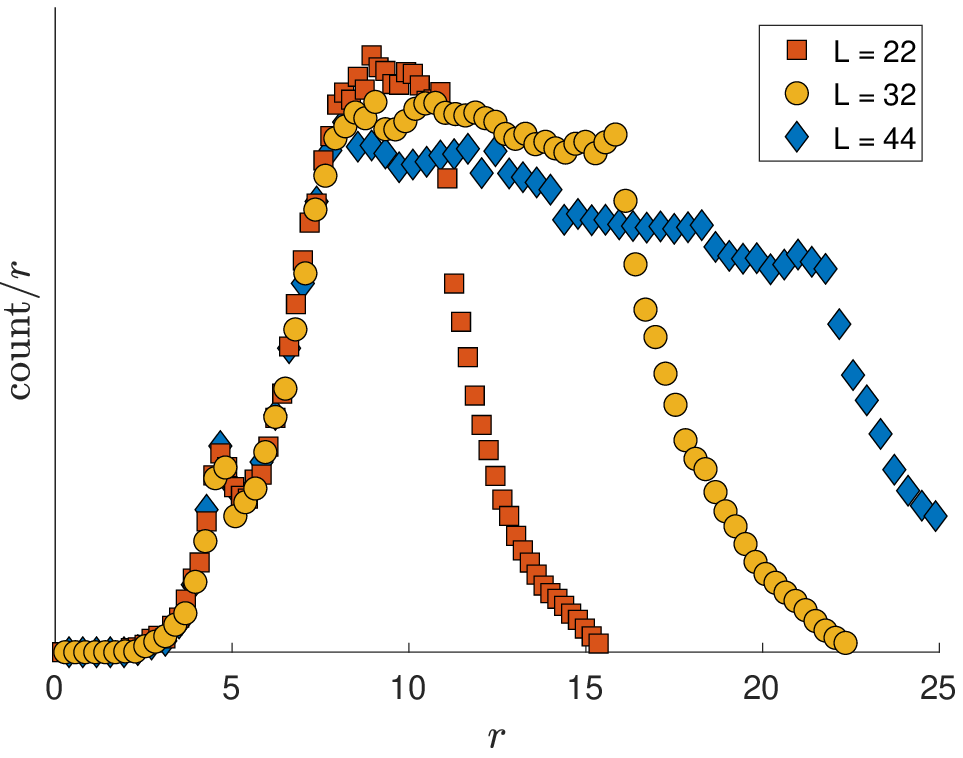}
    \caption{Counts for the occurrence of finding two rotors at a distance $r$ during a simulation for 3 different box sizes. The dimers are with spherical particles of radius 1 placed at a distance 6 (total length 8). The box is square with periodic boundary conditions and its size is 22, 30 and 44. The counts are represented with a normalization by $r$.}
    \label{fig:vary_L}
\end{figure}

For the result of figure \ref{fig:model_dumbbell}, we have used a box size of 22 space units (1 space unit is the radius of the two spheres of the dumbbell). To rule out any effect of the size of the box on the effective interaction potential between two rotating particles, we have simulated the particles dynamics for $L_b = 22$, $L_b=32$ and $L_b=44$ and the results are presented in figure \ref{fig:vary_L}. We observe that the interaction part of the potential is independent of the size of the box. 

\subsection{Dimensionless number $\omega_1/D_\theta$}
\begin{figure}
    \centering
    \includegraphics[width=0.5\linewidth]{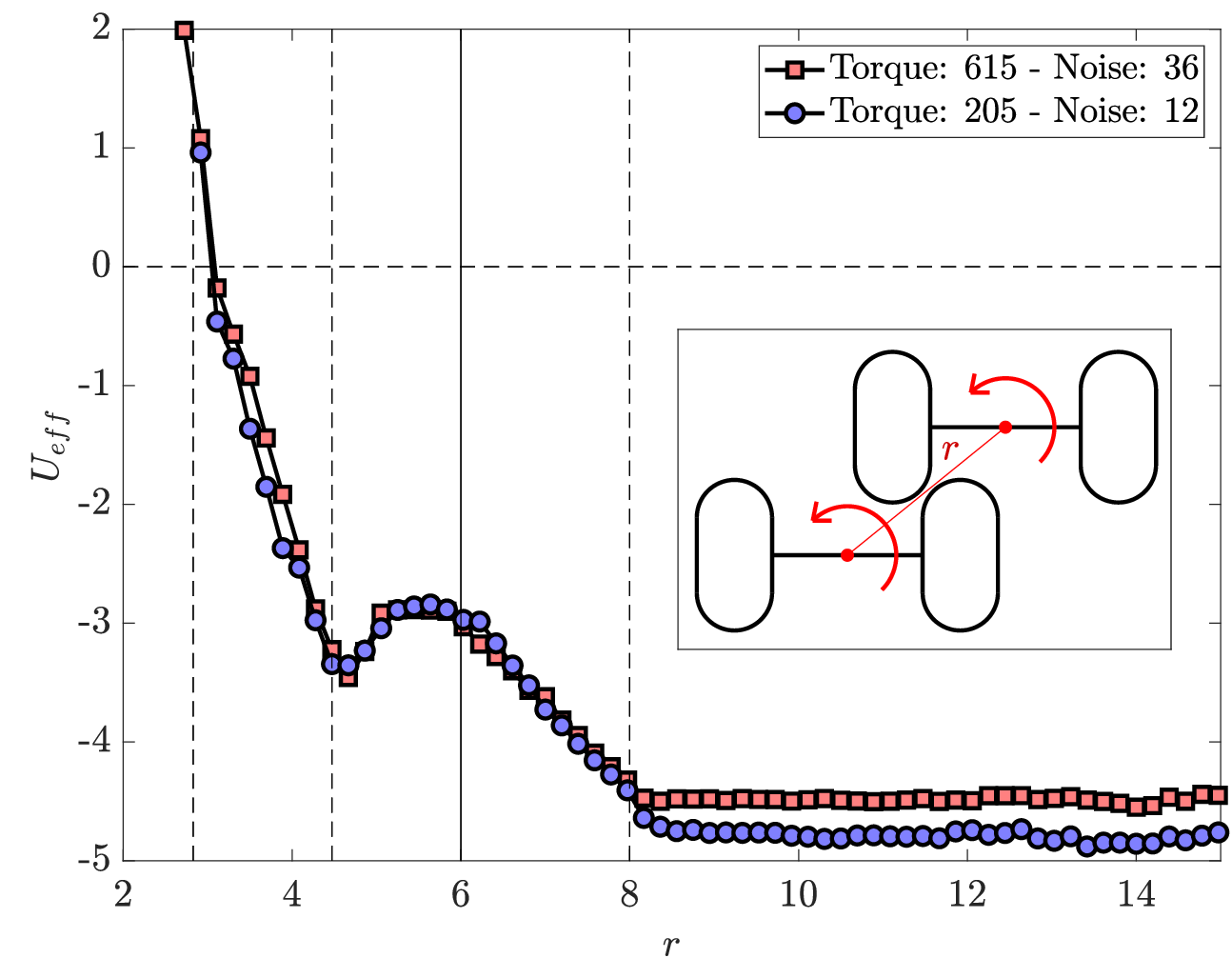}
    \caption{Simulation of the effective potential for the same ratio $\omega_1/D_\theta = 3.1$. There is a factor 3 in the torque ($\propto \omega_1$) and noise parameter ($\propto D_{\theta}$).  }
    \label{fig:compa_ratio}
\end{figure}

In figure \ref{fig:pot_energy}, we probe the effective potential evolution with the dimensionless number $\omega_1/D_\theta$ that represents the rotation activity compared to the Brownian noise via the angular diffusion coefficient $D_\theta$. In figure \ref{fig:pot_energy}, we showed the strengthening of the attractive potential by varying $\omega_1$ for a fixed $D_\theta$. To validate the relevance of the dimensionless number $\omega_1/D_\theta$, we present in figure \ref{fig:compa_ratio}
the interaction potential simulated for two pairs of values $\omega_1$ and $D_\theta = $ that have the same ratio $\omega_1/D_\theta$. 

\subsection{Effective potential}\label{app:eff_pot}
The point of simulating two particles and not more is to investigate the pair potential between two rotors without the problem of many-body interactions and dense regimes that would lead for example to crystallization even with purely repulsive granular particles. 
To infer an interaction potential between two dumbbells from the trajectory distribution, we use Boltzmann approach in which the probability to find two particle between $r$ and $r+dr$ is 
$$
dP(r,r+dr) = e^{-\frac{U(r)}{kT}} f(r)  dr
$$
in which $U(r)$ is an effective interaction potential, $kT$ is the thermal noise and $f(r)$ is a function that accounts for the finite size of the box. 

\begin{figure}[h!]
    \centering
    \includegraphics[width=0.5\textwidth]{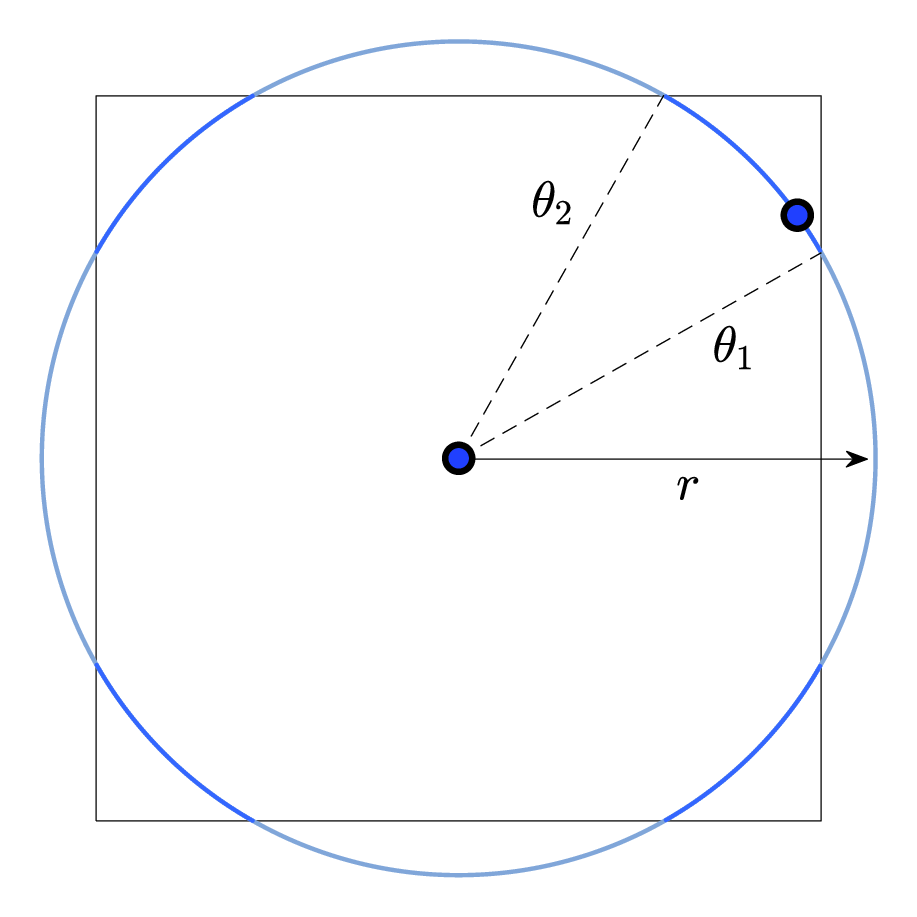}
    \caption{Representation of two particles randomly placed in a square box with periodic boundary conditions. Because of the periodic boundary conditions, the statistical properties of the two-particle interaction can be considered by shifting the particles positions such that one particle is always at the center of the box.  }
    \label{fig:part_box}
\end{figure}

To estimate $f(r)$, we solve the distance distribution between two point particles evenly distributed in a  box. In figure \ref{fig:part_box} we represent two particles in a periodic square box of size $L_b$. We used the periodicity of the box to place one of the particles at its center. 

If $r<L_b/2$
$$
f(r) = 2 \pi r
$$

If $r>L_b/2$
$$
f(r) = 4 l
$$
with $l$ the arc length of a circle of radius $r$ contained inside the box (represented by a thick line in figure \ref{fig:part_box}).
We have the relations: $\cos \theta_1 = L_b/2r$, $\sin \theta_2 = L_b/2r$ and $l = r(\theta_2-\theta_1)$.
which leads to
$$
f(r) = 4r\left(\arcsin\left(\frac{L_b}{2r}\right)-\arccos\left(\frac{L_b}{2r}\right) \right)
$$

\begin{figure}[h!]
    \centering
    \includegraphics[width=0.5\linewidth]{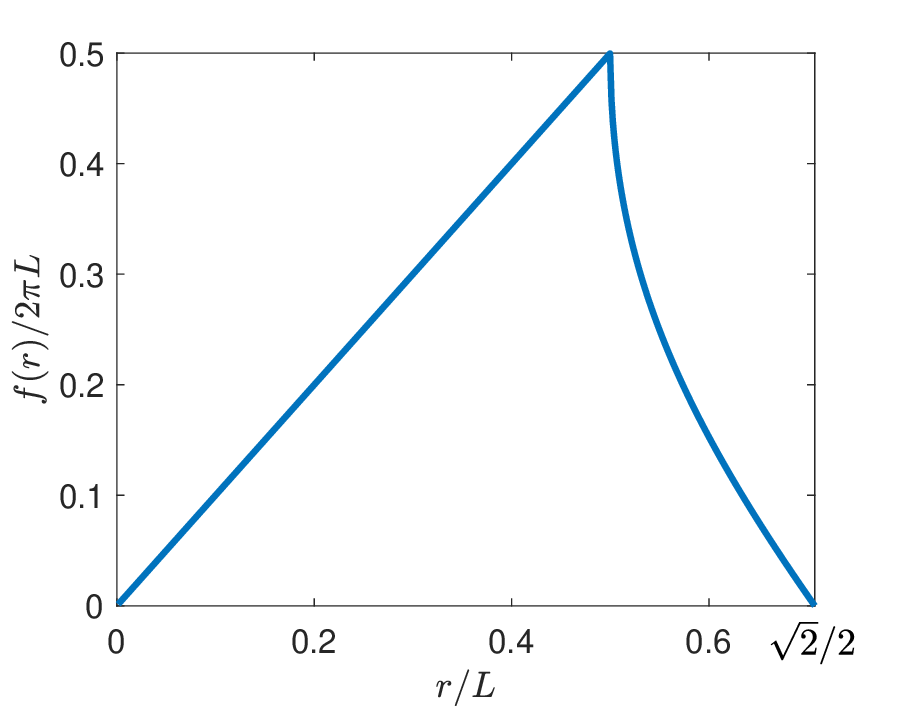}
    \caption{Probability to find two point-like particles (\textit{i.e.} non-interacting) at a distance $r$ in a periodic box of size $L_b$. This function $f(r)$ is used as a normalizing probability distribution in the case of interacting particles.}
    \label{fig:function_fr}
\end{figure}

Figure \ref{fig:function_fr} represents the function $f(r)/2\pi$ that is used to normalize the distance probability distribution in figure \ref{fig:brownian_dimers_map}.

To obtain the potential $U(d)$, we sum the probability of presence in figure \ref{fig:brownian_dimers_map} over $\varphi$ to have a probability function $P(d)$ and we can invert the Boltzmann relation:
$$
U(r) = -kT \log\left(\frac{1}{f(r)}\frac{dP(r)}{dr} \right)
$$

\subsection{Analytical predictions}
\subsubsection{Interaction potential in the Brownian limit for aligned and perpendicular sphere dumbbells}\label{app:pot_brownian}
In the Brownian limit, ie no rotational activity, the position distribution of two dumbbells is a steric problem. We can establish an analytical prediction for the dumbbell position distribution if we consider the two dumbbells aligned ($\varphi = 0$) or perpendicular ($\varphi=\pi/2$).

\begin{figure}[h!]
    \centering
    \includegraphics[width=.5\linewidth]{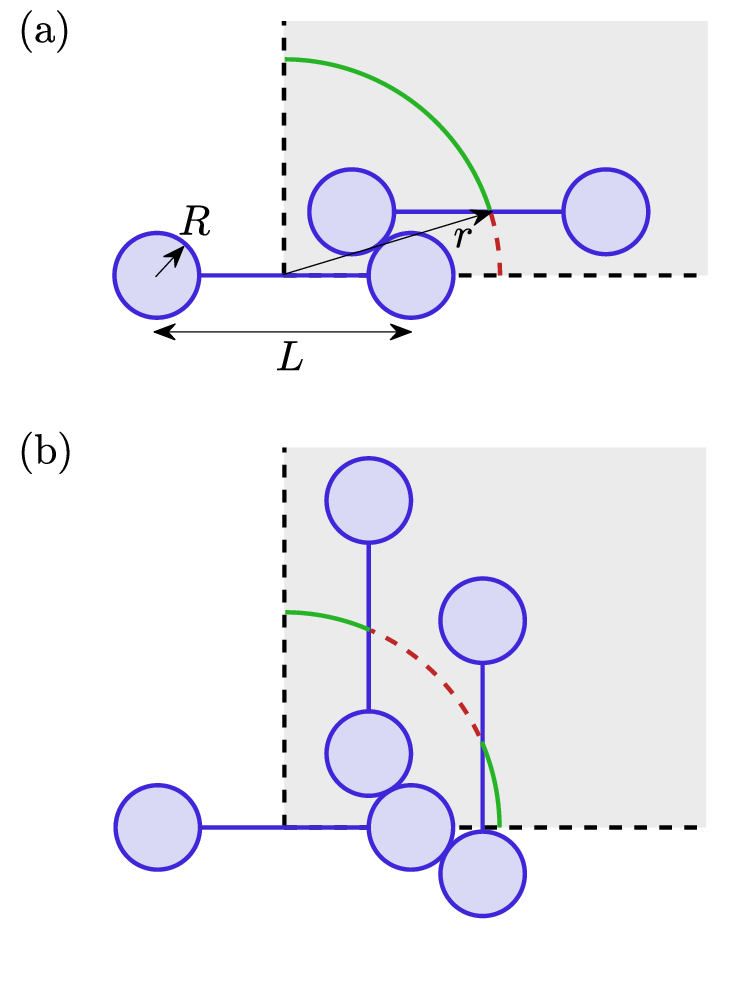}
    \caption{a) Representation of the limit angle of contact for two aligned dimers at a distance $r<L+2R$. b) Representation of the limit angle of contact for two perpendicular dimers at a distance $r<L+2R$. }
    \label{fig:two_dumb}
\end{figure}

In figure \ref{fig:two_dumb}, we represent the accessible configurations for two aligned dumbbells at a distance $r$. 
For the arc circle in the upper right quadrant, the accessible configurations correspond to the part of the circle with a solid line and the forbidden configurations are represented by a dashed line. 

The probability density $\mathcal{P}_\parallel = dP_\parallel(r,r+dr)/dr$, with $dP_\parallel(r,r+dr)$ the probability to find a dumbbell with the same orientation at a distance between $r$ and $r+dr$ is given by:
\begin{eqnarray}
0<r<{r_\parallel}^{min},\hspace{1cm}  && \mathcal{P}_\parallel = 0\\
{r_\parallel}^{min}<r<{r_\parallel}^{bound},\hspace{1cm}  && \mathcal{P}_\parallel = 2\pi r  - 4 r\arcsin\left(\frac{R}{r} \right)\\
{r_\parallel}^{bound}<r<{r_\parallel}^{rep},\hspace{1cm}  && \mathcal{P}_\parallel   = 2\pi r  - 4r \arcsin\left(\frac{\sqrt{4R^2-\left(\frac{r^2 - d^2 - 4R^2 }{2d}\right)^2}}{r}\right)\\
{r_\parallel}^{rep}<r< \frac{\sqrt{2}}{2}L_b,\hspace{1cm}  && \mathcal{P}_\parallel = 2\pi f(r) r 
\end{eqnarray}
with ${r_\parallel}^{min} = 2R$, ${r_\parallel}^{bound} = \sqrt{(d-\sqrt{3}R)^2 + R^2}$ and ${r_\parallel}^{rep} = d + 2R$

For two dumbbells with a perpendicular orientation, we have:
\begin{eqnarray}    
0<r<{r_\perp}^{min},\hspace{1cm}  && \mathcal{P}_\perp = 0\\
{r_\perp}^{min}<r< {r_\perp}^{bound},\hspace{1cm}  && \mathcal{P}_\perp =4\pi r - 8 r\arcsin\left(\frac{d + 2R}{2r}\right)\\
{r_\perp}^{bound} <r<{r_\perp}^{rep},\hspace{1cm}  && \mathcal{P}_\perp =  4\pi r - 8r \arcsin\left(\frac{\frac{d}{2} + 2R\sin(\alpha)}{r}\right)\\
{r_\perp}^{rep}<r<\frac{\sqrt{2}}{2}L,\hspace{1cm}  && \mathcal{P}_\perp  = 2 \pi f(r) r 
\end{eqnarray}
with $\alpha = \arcsin\left(\frac{ 2r^2 - L^2 - 8 R^2}{4\sqrt{2}RL}\right) + \frac{\pi}{4}$, ${r_\perp}^{min} = \frac{L}{2}+R$, ${r_\perp}^{bound}=\sqrt{\left(\frac{L}{2} - \sqrt{3}R\right)^2 + \left(\frac{L}{2}+R\right)^2}$ and ${r_\perp}^{rep}=\frac{\sqrt{2}}{2} L+2 R$

\begin{figure}[h!]
    \centering
    \includegraphics[width=0.5\linewidth]{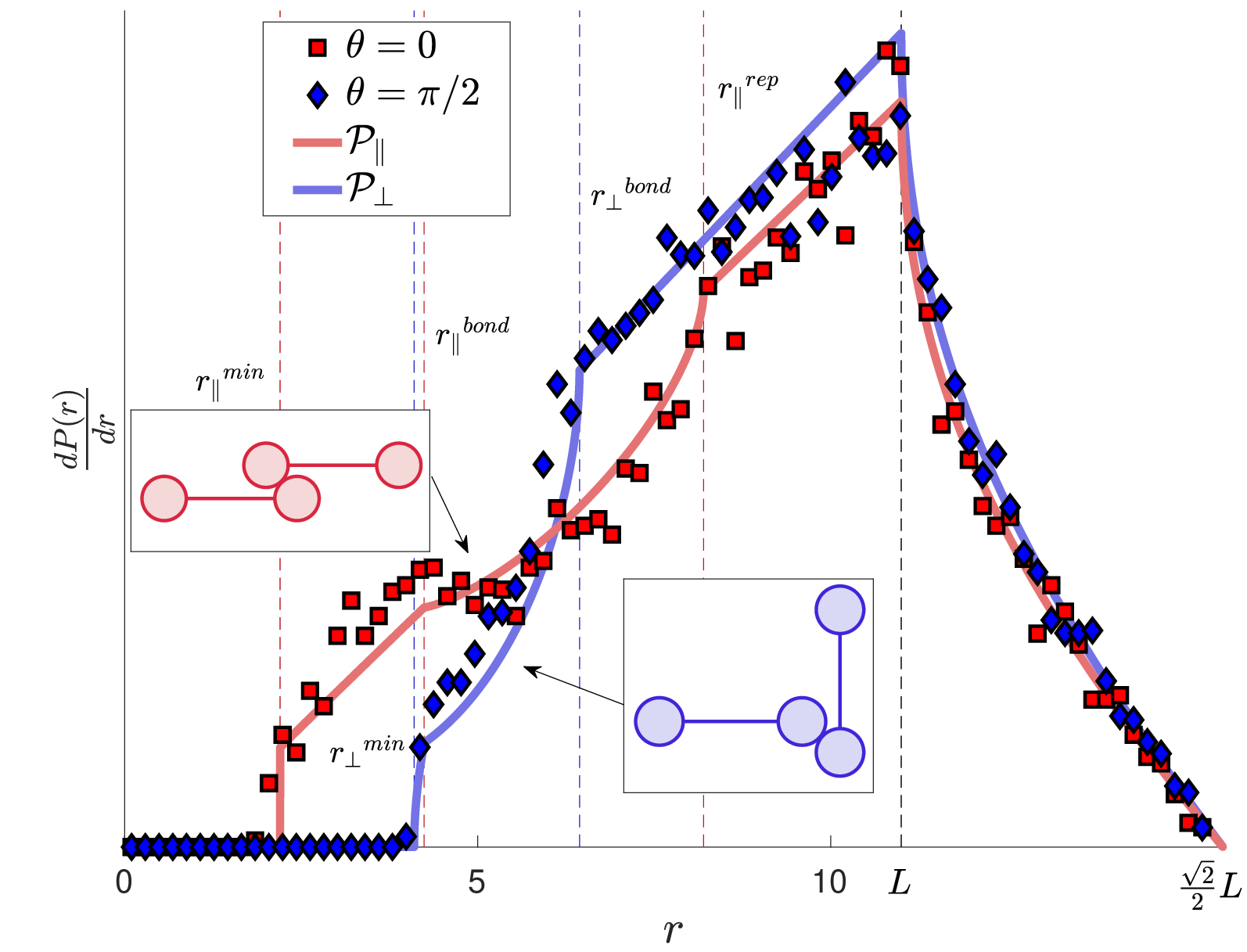}
    \caption{Probability of finding two parallel and two perpendicular dimers at a distance $r$. The solid lines are analytical results based on the dimer geometry. The data points are obtained by numerical simulation of Brownian dimers (no rotation).}
    \label{fig:P_para_perp}
\end{figure}

The functions $\mathcal{P}_\parallel$ and $\mathcal{P}_\perp$ are represented in figure \ref{fig:P_para_perp} by solid lines. The square and diamond symbols are the probability distributions from the particle simulation from the contour plot in figure \ref{fig:model_dumbbell} b in which we select the horizontal lines $\varphi = 0$ and $\varphi = \pi/2$ (bin width 0.09 rad).

\begin{figure}[h!]
    \centering
    \includegraphics[width=0.5\linewidth]{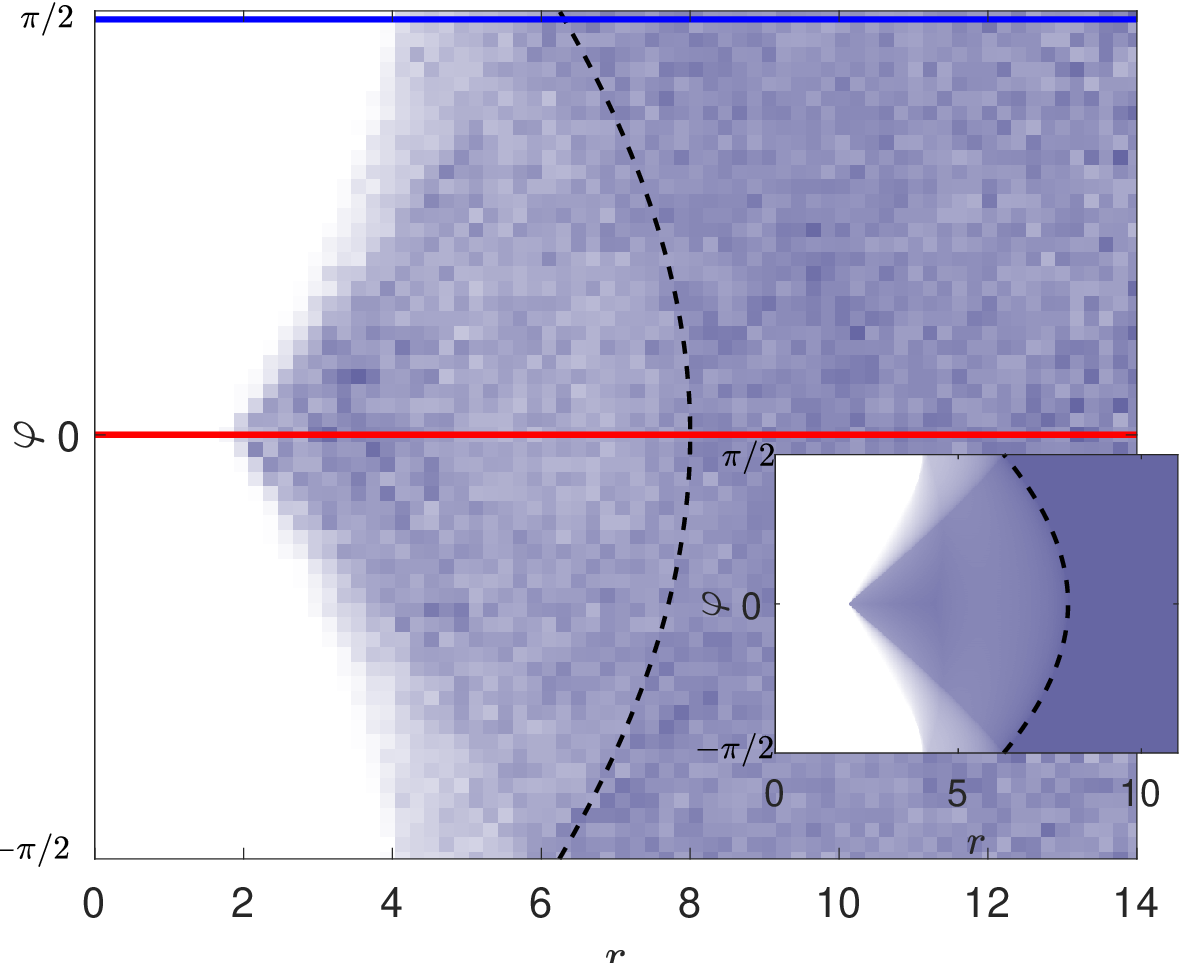}
    \caption{Representation of the probability of finding two dimers at a distance $r$ with an orientation difference $\varphi$. The map is obtained by the analysis of long trajectories from the simulation of two Brownian dimers. The inset is an analytical result constructed by uniformly sampling all the possible configurations $\{r,\theta, \varphi\}$ for which the two dimers do not overlap. }
    \label{fig:distr_no_torque_sphere}
\end{figure}

In figure \ref{fig:distr_no_torque_sphere}, we represent the distribution of position for the two sphere dumbbells in the space $\{r,\varphi \}$ in the Brownian regime. The horizontal lines at $\varphi = 0$ and $\varphi=\pi/2$ corresponds to the two plots in figure \ref{fig:P_para_perp}. We use two distinct methods. The main figure is the probability distribution obtained from the simulated trajectory of two Brownian dumbbells in a periodic box. The contour in the inset is the probability of presence obtained by numerically generating all the possible configuration $\{ r, \theta, \varphi\}$ and averaging over $\varphi$ by counting the fraction of non-overlapping configurations.
Here, we are in the Brownian regime and the two approaches match because of ergodicity. For active regimes, only the method with the simulated trajectory is valid to estimate the probability distribution of the dumbbells.

\subsubsection{Interaction potential in the Brownian limit for aligned and perpendicular spherocylinder dumbbells}
%In figure 2 of the main article, we use dumbbells made of spherocylinders. Figure \ref{fig:schema_spherocyl_dimers} represents the structure of the dumbbell with a set of $2\times 7$ spheres with $R = 1$ used to form two spherocylinders for each dumbbells. 
%\begin{figure}[h!]
%    \centering
%    \includegraphics[width=0.5\linewidth]{schema_spherocyl_dimers.png}
%    \caption{Representation of the contact of two dimers made of spherocylinders. }
%    \label{fig:schema_spherocyl_dimers}
%\end{figure}
 
\begin{figure}[h!]
    \centering
    \includegraphics[width=0.5\linewidth]{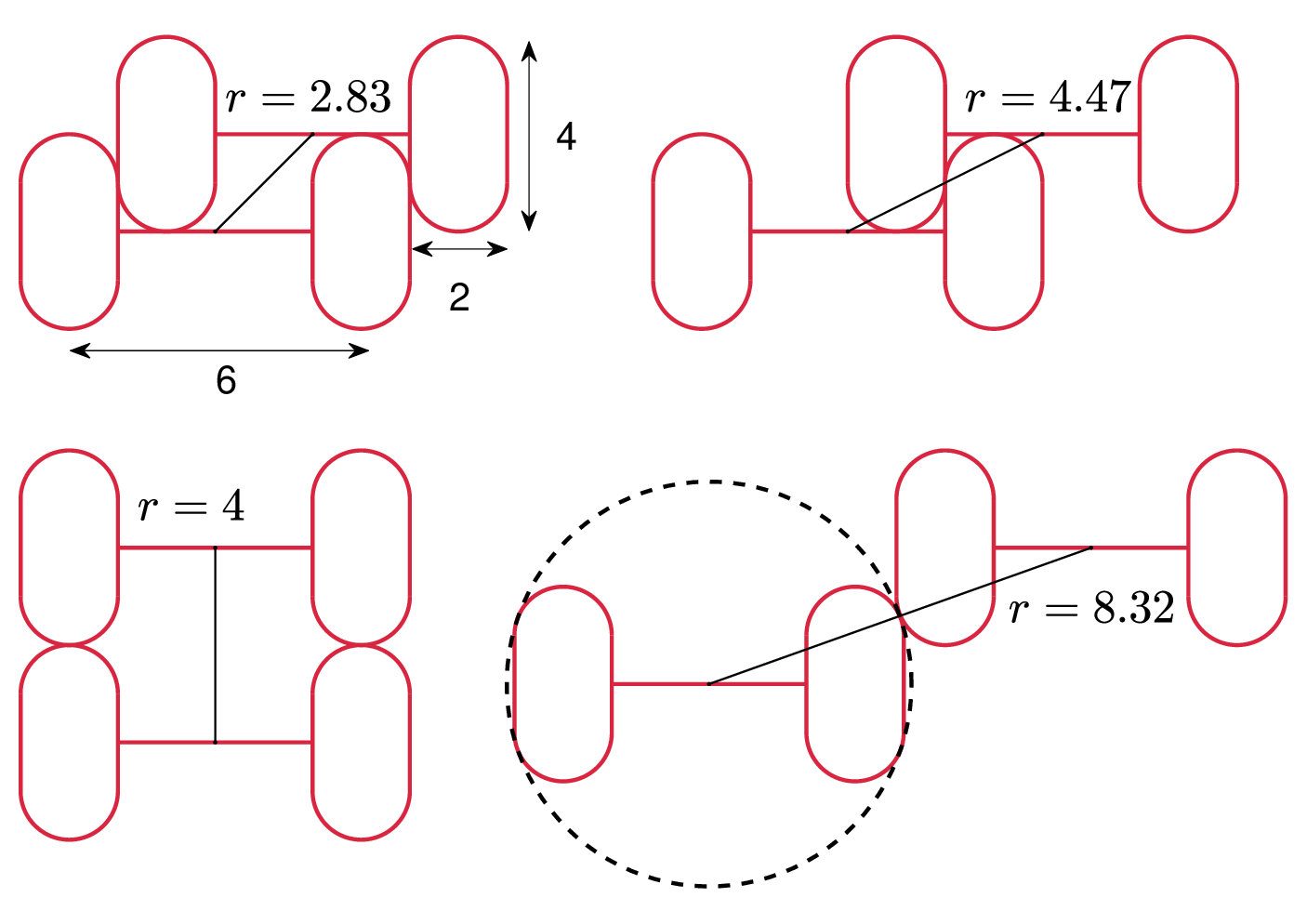}
    \caption{Representation of possible contact configurations of two dimers with spherocylinders.}
    \label{fig:spherocyl_configs_brownian}
\end{figure}

\begin{figure}[h!]
    \centering
    \includegraphics[width=0.5\linewidth]{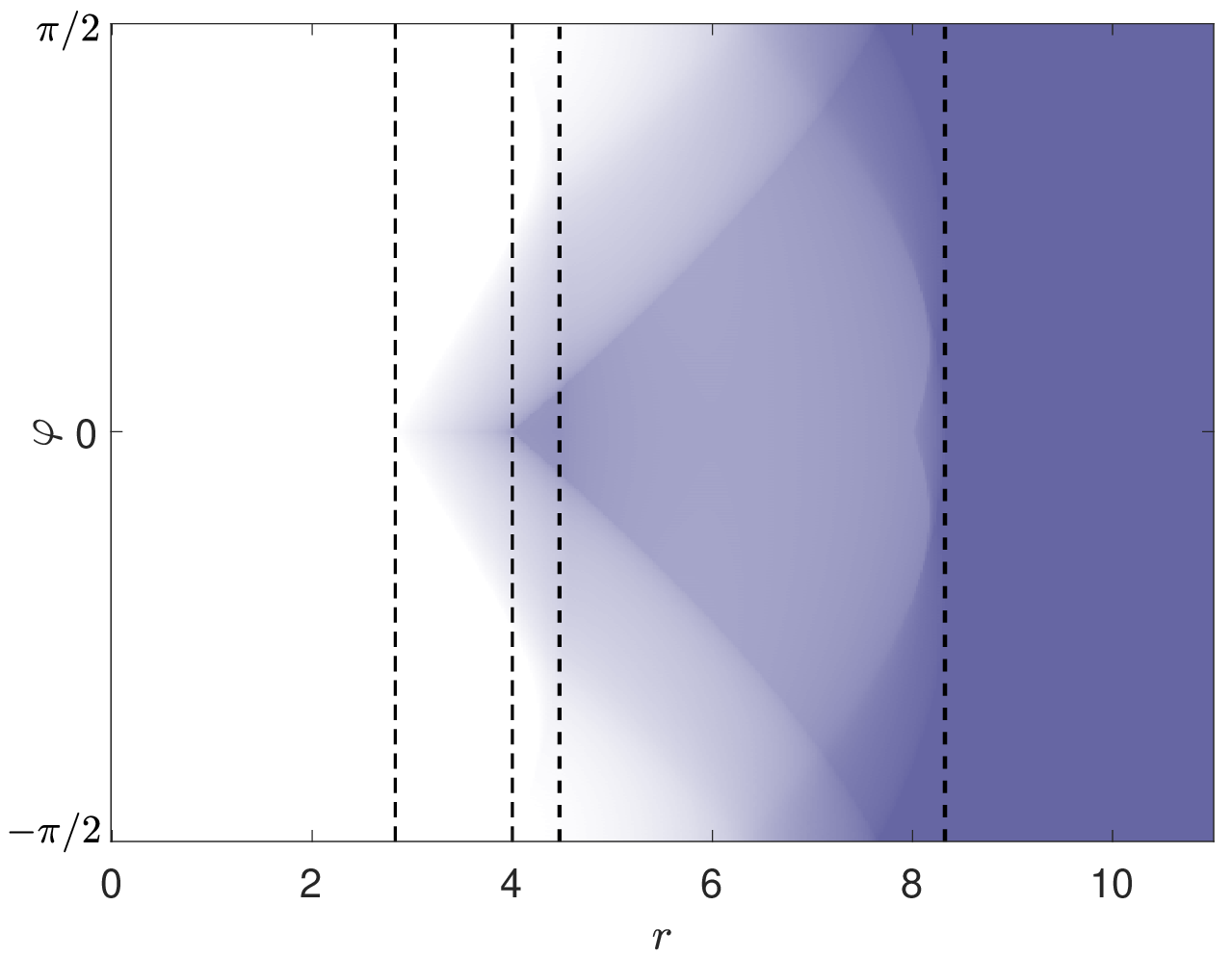}
    \caption{Probability of finding two dimers at a distance $r$ and an orientation difference $\varphi$. The dimers are made of two spherocylinders as it is in figure \ref{fig:spherocyl_configs_brownian}. The map is obtained by generating all the configurations $\{r, \theta, \varphi\}$ with uniform sampling and by conserving all the non-overlapping configurations, after an averaging over $\theta$.}
    \label{fig:distr_spherocyl_brownian}
\end{figure}

In figure \ref{fig:distr_spherocyl_brownian}, we represent the position distribution as a function of the distance $r$ and the relative orientation $\varphi$ of the two dumbbells. We identify 4 distances $r$ corresponding to the situations depicted in figure \ref{fig:spherocyl_configs_brownian}.

\begin{figure}[h!]
    \centering
    \includegraphics[width=0.45\linewidth]{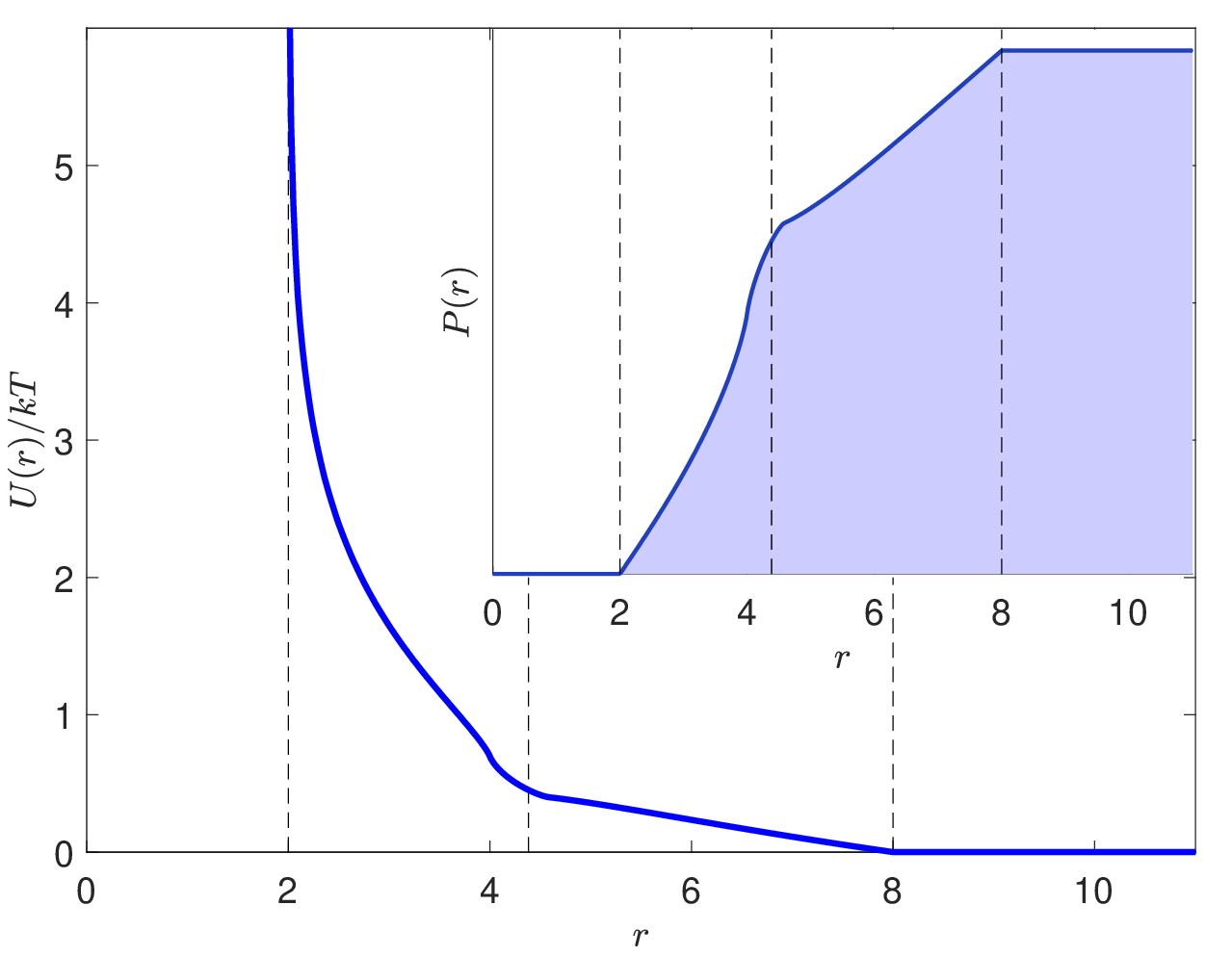}
    \caption{Construction of the interaction potential function $U(r)$ in the Brownian case (no rotational activity, $\omega_1/D_\theta=0$). The inserted plot $P(r)$ is after averaging on $\varphi$ the plot in figure \ref{fig:distr_spherocyl_brownian}. The main plot is the Brownian limit for the interaction potential $-\log P(r)$ with $\omega_1 = 0$.}
    \label{fig:distr_spherocyl_brownian_2}
\end{figure}

In figure \ref{fig:distr_spherocyl_brownian_2}, we represent the probability $P(r)$ obtained by averaging the contour map over $\varphi$ and the Boltzmann potential $U_{Brownian}/kT = -\log(P(r))$ with $\omega_1 = 0$ (Brownian limit). This potential $U_{Brownian}$ is the one used in figure \ref{fig:pot_energy} for $\omega_1/D_\theta = 0.0$.

\subsection{Chain statistics}\label{app:chain_stat}
\begin{figure}[h!]
    \centering
    \includegraphics[width=0.5\linewidth]{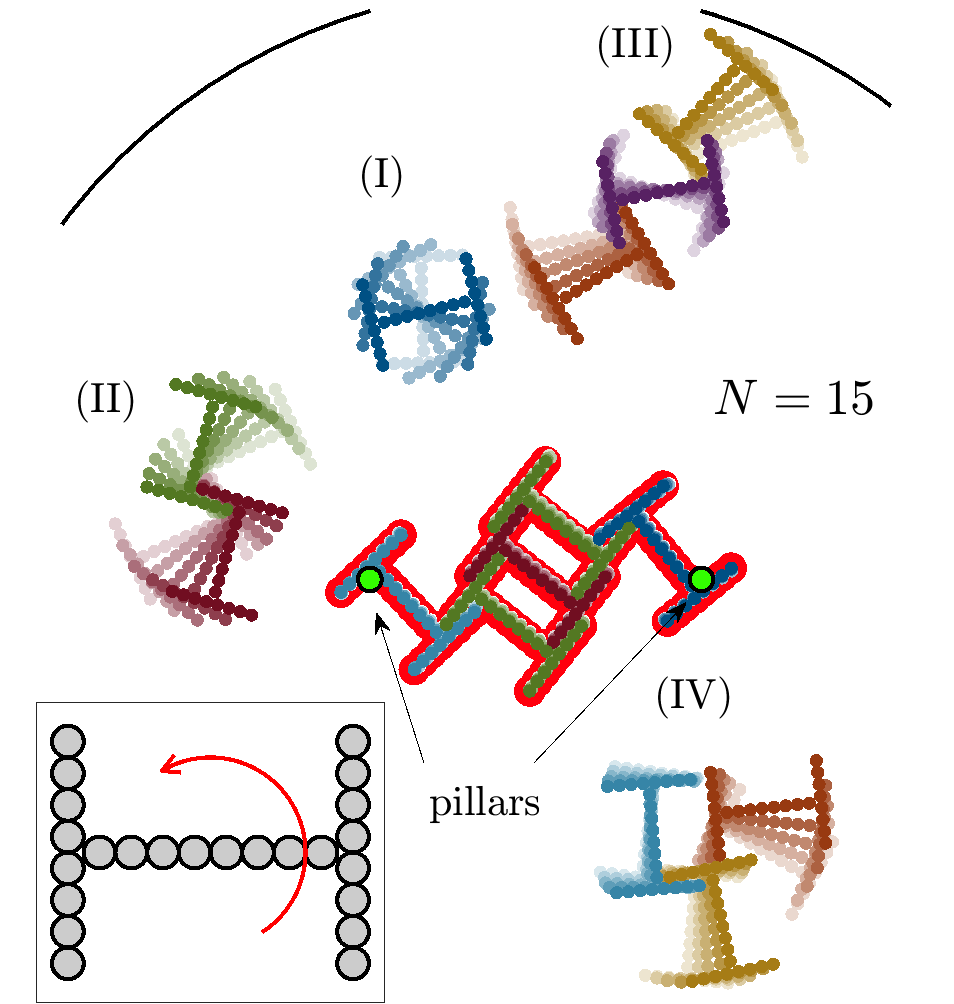}
    \caption{View of the simulation of 15 H-like rotors in a circular box. Diameter 100, distance between the pillars $d=28$.}
    \label{fig:visu_simu}
\end{figure}

In figure \ref{fig:simu_chains}, we present a numerical study of the waiting time associated with the formation and dissolution of $N$ rotors into chains. A snapshot for one of the numerical simulations with $N=15$ rotors is represented in figure \ref{fig:visu_simu} (the rotors are labelled $11\times9$ to recall that they occupy an horizontal distance of 11 spheres and a vertical distance of 8 spheres). There are 4 groups of rotors with (I) a single rotor, (II) a steady structure of two rotors, (III) a slowly evolving structure of 3 rotors and (IV) a steady chain of 5 rotors linking two static anchoring points at a distance $r = 32$. 

The simulation is run in a circular domain of diameter 200 for a duration corresponding to 347 rotation times of a free rotor. We evaluate the presence of a chain by determining if there is a set of connected rotors in contact with the two anchoring points separated by a distance $d$. If a chain is present at the end of the simulation, we set the end of the simulation time as the dissolving of the chain (this tends to underestimate the waiting time for dissolution).
%\subsection{Other geometries}

\section{Mechanical model of the cohesive interaction}\label{app:model_interaction}
\subsection{Presentation of the model}
\begin{figure}[h!]
    \centering
    \includegraphics[width=0.5\linewidth]{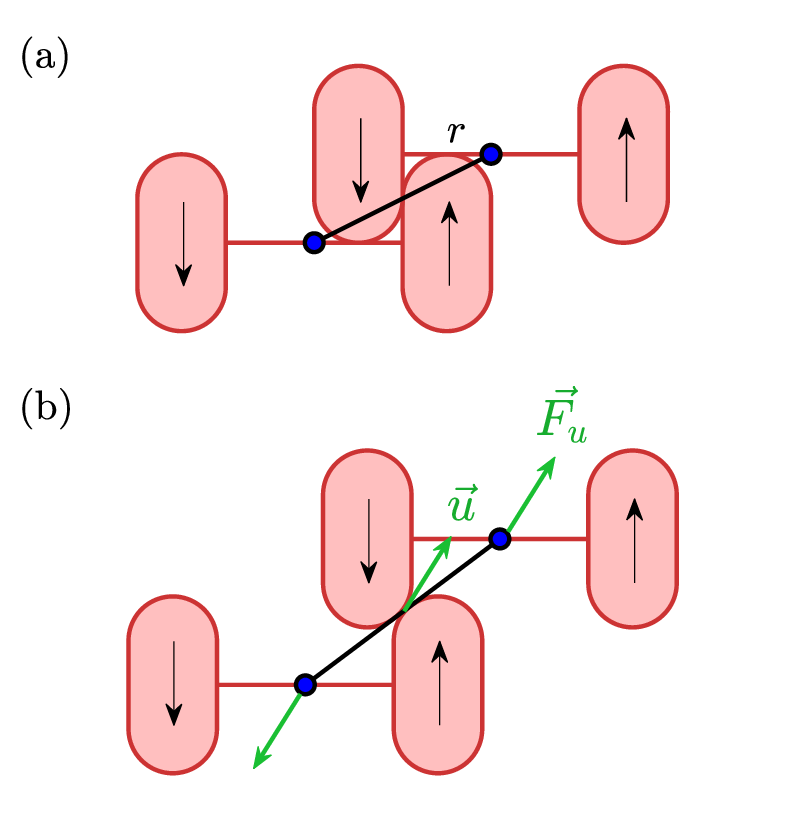}
    \caption{a) Representation of two dumbbells made of spherocylinders at the bound configuration ($r = \sqrt{4^2 + 2^2}\approx 4.47$). b)  Representation of two dumbbells in the attractive range ($r\approx 5.1$).  }
    \label{fig:mod_pres}
\end{figure}

The model aims at estimating the interaction potential between two dumbbells under the effect of rotational activity. The bound state  represented in figure \ref{fig:mod_pres}a is the reference state at mechanical equilibrium. In the example of figure \ref{fig:mod_pres}b, without the additional forces $F_u$, the dumbbells would come back to the configuration of figure \ref{fig:mod_pres}a under the action of the torque. To maintain the dumbbells in the configuration of figure \ref{fig:mod_pres}b, an additional pair of forces $F_u$ is required. $u$ is a unit vector that is tangent to the contact of the two dumbbells and $F_u$ is parallel to $u$. 
the external forces are pulling on the two dumbbells which is consistent with the rotational activity that would get the dumbbell back the configuration of the bound state in figure \ref{fig:mod_pres}a.

The force $F_u$ can be computed for each contact configuration of the two dumbbells (see the following section). From $F_u$, we can estimate the interaction potential by integration $U_{Active} \sim \int F_u dr$.

%\clearpage

\subsection{Derivation of the active component of the interaction}\label{sec:deriv_F}
The interaction potential between two dumbbells is of the form 
\begin{equation}
    U_{eff} = U_{Brownian}(r) + U_{Active}(r)\label{eq:model}
\end{equation}
in which $U_{Brownian}$ is the potential in the absence of activity and $U_{active}$ is the contribution of the activity on the pair interaction for the two dumbbells.

To compute $U_{active}$, we use a mechanical model with the geometry presented in figure \ref{fig:mod_pres}.
We consider that the dumbbells are frictionless objects submitted to a pair of active forces $F_A$ and $-F_A$ and to a pair of linear viscous  forces $\vec{F_{visc}} = -\alpha \vec{v}$ with $v = r \omega$ the velocity of each shperocylinders at distance $r$ of the geometric center and $\alpha$ a viscous coefficient. $\omega$ is a free parameter of the model, set by the balance between active torque and viscous torque. 
We consider that the two dumbbells are parallel ($\varphi = 0$). 
\begin{figure}[h!]
    \centering
    \includegraphics[width=0.5\linewidth]{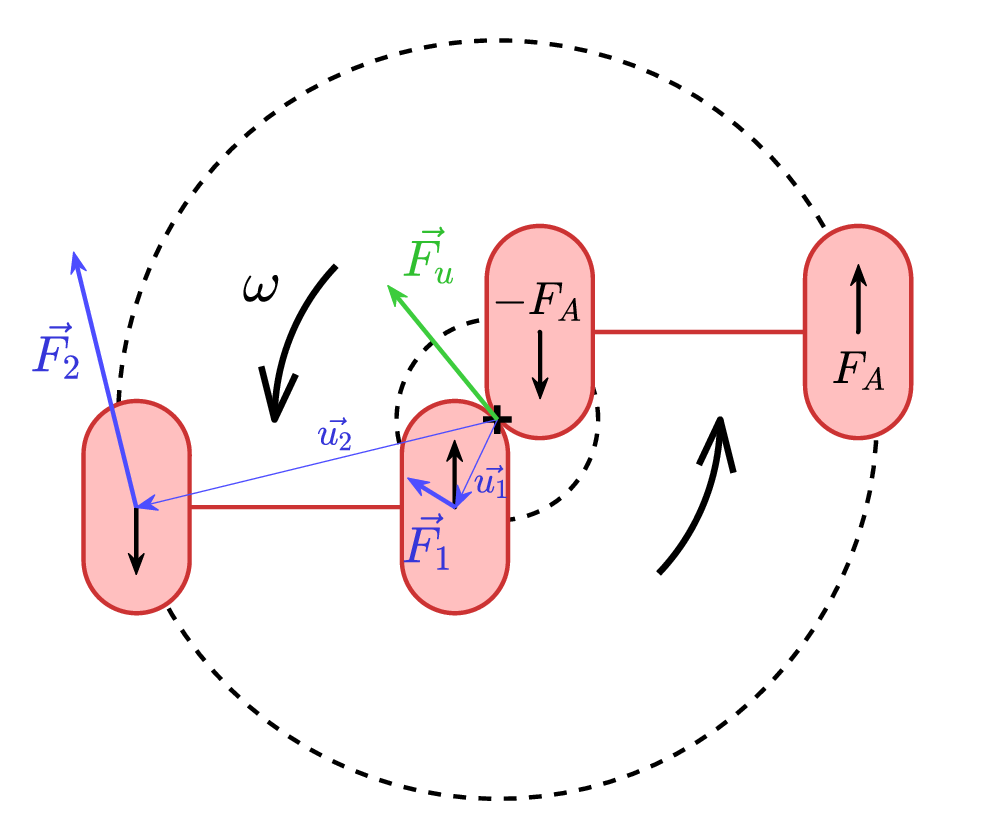}
   \caption{Representation of two frictionless dimers at contact in solid body rotation. }
  \label{fig:mech_model}
\end{figure}

To estimate the potential, we compute the additional force $F_u$ needed to maintain a pair of dumbbells in synchronous rotation under the equilibrium of active forces and viscous friction. For a configuration such as the one presented in figure \ref{fig:mech_model}, we can deduce from the particles geometry the vectors $\vec{u_1}$ and $\vec{u_2}$ that points between the contact point in the center and the position of the two spherocylinders. If we use the unit vector $\vec{u} = (\sin \beta, -\cos \beta)$ we have:
\begin{eqnarray}
    \vec{u_1} &=& \left(\begin{array}{l}-R \cos \beta\\ - (R  +R\sin \beta)\end{array}\right)\\
    \vec{u_2} &=& \left(\begin{array}{l}-(L+R \cos \beta)\\ - (R  +R\sin \beta)\end{array}\right)
\end{eqnarray}
with $r = \|\vec{u_1}+\vec{u_2}\|$ with $r$ the distance between the two dumbbells centers.

We assume that the pair of dumbbells is in solid rotation which means that the velocity of the spherocylinder elements is $\vec{u_i}\times \vec{\omega}$ with $\omega$ the angular velocity of the structure. The forces $F_1$ and $F_2$ are viscous forces. $F_1$ and $F_2$ are tangent to the rotation (oriented perpendicularly to $\vec{u_1}$ and $\vec{u_2}$ respectively) and their magnitude is $\alpha \omega \|\vec{u_1}\|$ and $\alpha \omega\|\vec{u_2}\|$ respectively, with $\alpha$ the viscous friction coefficient. The amplitude for the associated moment for the viscous forces is $\alpha \omega \|\vec{u_1}\|^2$ + $\alpha \omega \|\vec{u_2}\|^2$.
We have the relations
\begin{eqnarray}
    \| \vec{u_1} + \vec{u_2}\|^2 &=& r^2\label{eq:model_r2}\\
\| \vec{u_1} - \vec{u_2}\|^2 &=& L^2\\
\| \vec{u_1} \|^2 + \|\vec{u_2}\|^2 &=& \frac{1}{2}\left[\| \vec{u_1} + \vec{u_2}\|^2 +  \| \vec{u_1} - \vec{u_2}\|^2\right]
\end{eqnarray}

The angular frequency is given by the balance between the torque from the pair of active forces and the viscous friction:
\begin{equation}
    \alpha\omega \left(\frac{r^2}{2} + \frac{L^2}{2} \right) = F_A L\label{eq:model_part1}
\end{equation}
with $F_A$ the magnitude of the pair of forces for the active torque $\Gamma =  F_A L$.

For a contact configuration, there is no force balance for the solid body rotation unless a tangential force $F_u$ is added parallel to the tangent vector $\vec{u}$. The magnitude of this force $F_u$ equals the sum of the projected forces $F_1$ and $F_2$ on the tangent vector $\vec{u}$. $F_1$ and $F_2$ equal the viscous forces acting on the spherocylinders with $\vec{F_i} = \alpha  \vec{u_i}\wedge  \omega\vec{u_z}$ for $i=1,2$.

We have:
\begin{eqnarray}
\vec{F_1}.\vec{u} &=& -\alpha \omega\left(R \sin \beta +R \sin^2 \beta +R \cos^2\beta\right) \label{eq:model_F1u} \\
\vec{F_2}.\vec{u} &=& -\alpha \omega\left(R \sin \beta +R \sin^2 \beta +L \cos \beta + R\cos^2 \beta\right) \label{eq:model_F2u}
\end{eqnarray}
and thus
\begin{equation}
 \vec{F_1}.\vec{u}+\vec{F_2}.\vec{u} = -2\alpha \omega \left(R\sin \beta + L \cos \beta + 2R \right) \label{eq:model_F.u} 
\end{equation}
We can use the relation (\ref{eq:model_r2}) with the expressions of $\vec{u_1}$ and $\vec{u_2}$ and we get:
\begin{equation}
r^2 = L^2 + 4R\left[L \cos \beta + 2R\sin \beta -2 R \right]\label{model_r2_L2}
\end{equation}

Combining equations (\ref{eq:model_part1}), (\ref{eq:model_F.u}) and (\ref{model_r2_L2}), we finally obtain:
\begin{equation}
   F_u = \vec{F_1}.\vec{u}+\vec{F_2}.\vec{u}= -2\frac{F_A L}{r^2 + L^2} \left( \frac{r^2 - L^2 }{4R}\right),\hspace{.5cm} 4.47<r<8.32\label{eq:model_part2}
\end{equation}

\begin{figure}[h!]
    \centering
    \includegraphics[width=0.5\linewidth]{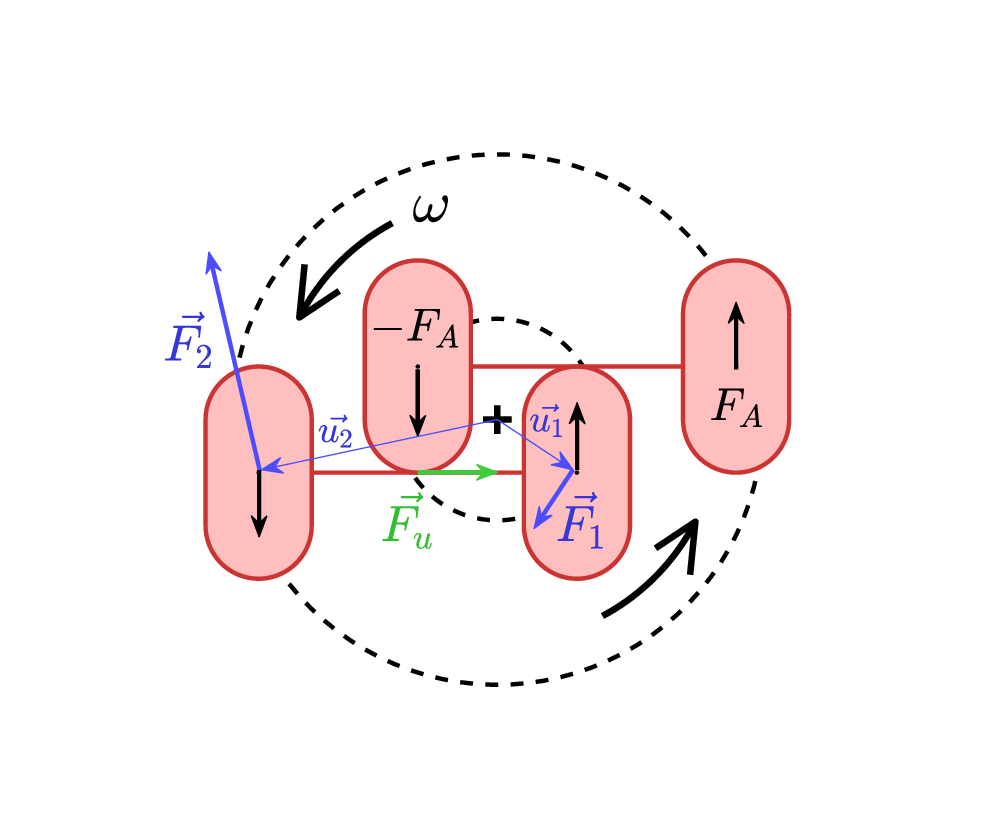}
    \caption{Contact configuration for close dimers ($2.83<r<4.47$).}
    \label{fig:mech_model_2}
\end{figure}

In figure \ref{fig:mech_model_2}, we represent the configuration when the dumbbells are closer than the binding distance.

By using the same approach as in figure \ref{fig:mech_model}, we find that the contact force to maintain the solid body rotation is 
\begin{equation}
     F_u = -4R\frac{F_A L}{r^2 + L^2},\hspace{.5cm} 2.83<r<4.47
\end{equation}

\begin{figure}[h!]
    \centering
    \includegraphics[width=0.5\linewidth]{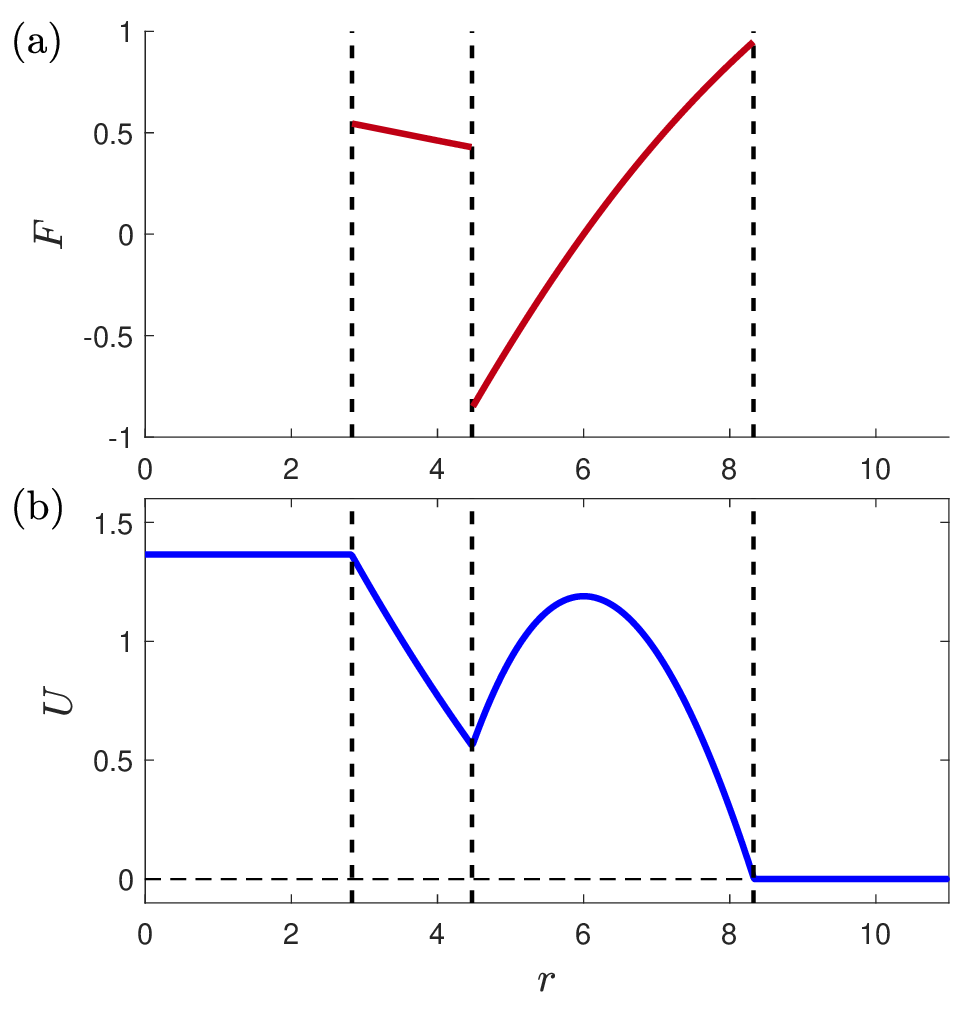}
    \caption{a) Additional force $F_u$ needed to maintain two frictionless dimers in solid body rotation as a function of $r$. b) Potential obtained by integration of the force $F_u$. }
    \label{fig:force_potential_spherocyl}
\end{figure}

To obtain the interaction potential $U_{Active}$ in figure \ref{fig:force_potential_spherocyl} , we use the scaling relation
$U_{Active} \sim -\int_\infty^r  \|\vec{F_u}\| dr  $ with the integration constant $U_0$ at $r = \infty$. 
%\clearpage

\subsection{Physical interpretation of the offset parameter in the model}\label{app:interpret}

\begin{figure}[h!]
    \centering
    \includegraphics[width=0.5\linewidth]{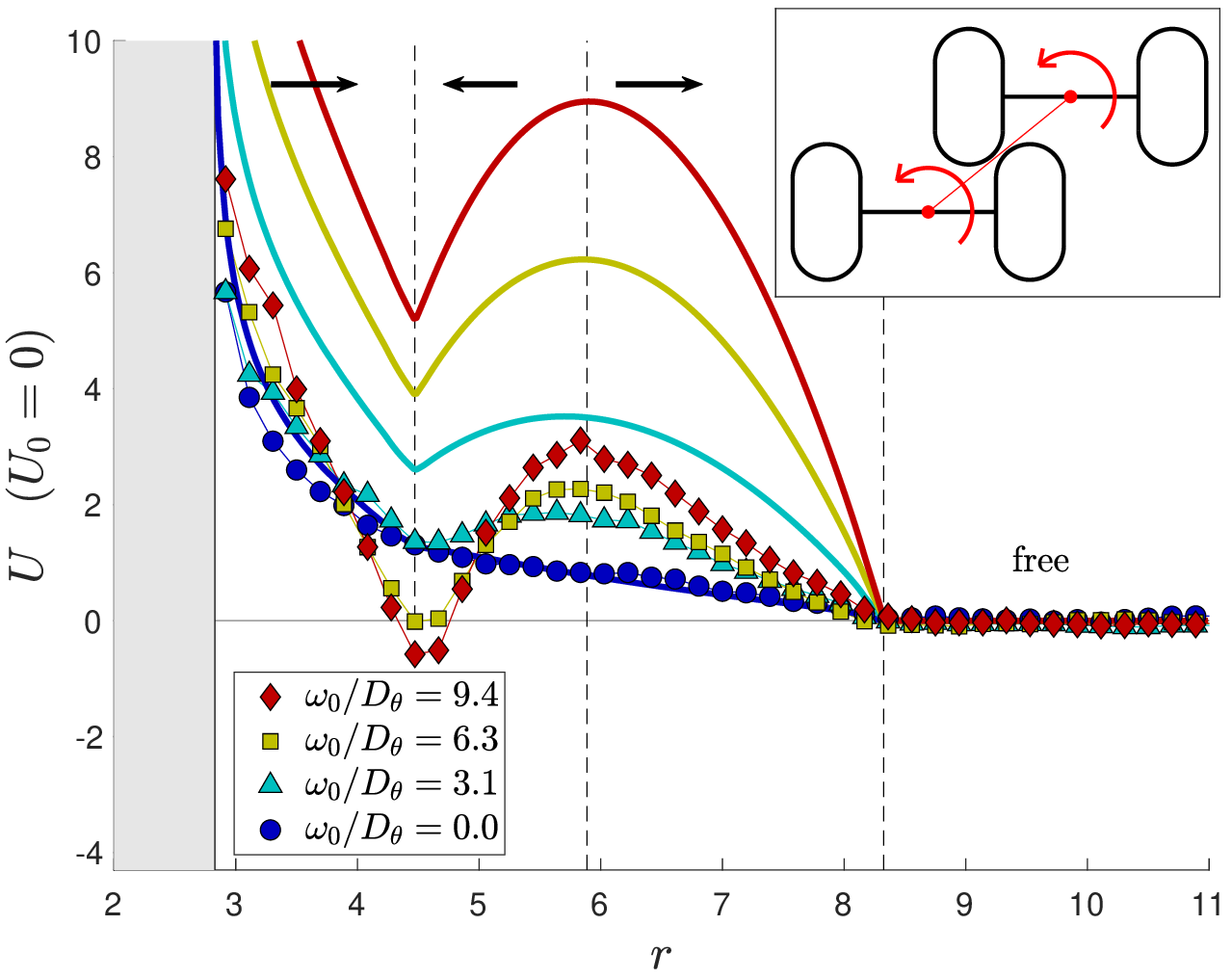}
    \caption{Effective potential as a function of distance (same data as in figure \ref{fig:pot_energy}). The model is without adjustable offset to show the inconsistency of the model when the free region is taken as a reference. }
    \label{fig:3_no_offset}
\end{figure}

To discuss the physical origin of the offset term $U_0$, we first represent the model without offset term in figure \ref{fig:3_no_offset}.

As shown in figure \ref{fig:3_no_offset}, the model does not correctly capture the connection between the free regime $U=0$ in which the particle are freely rotating without interacting and the interacting regime in which we assumed that the potential can be obtained by assuming a synchronous rotation of the two connected dimers which is not the case in the free regime. 

\section{Movies\cite{supp_movie}}

M1 

$N = 6$ isolated spinners in a circular box (diam: 38 cm) with a central pillar.

M2

$N = 6$ spinners in a circular box (diam: 38 cm) with a central pillar. Two spinners are paired on the right side and another spinners interacts with the pillar.

M3 

$N = 10$ spinners in a circular box (diam: 38 cm) with two pillars separated by a distance $d = 16$ cm. 

M4 

N = 4 spinners initially placed as a chain between two pillars separated by a distance d = 12 cm.

M5 

$N = 9$ spinners in a circular box (diam: 38 cm) with two pillars separated by a distance $d = 14$ cm. A chain of $N_{ch} = 5$ self-assembles and remains for the duration of the acquisition.

M6 

View of the molecular dynamics simulation with rotational activity (self-imposed torque). $r$ is the distance between the two spherical dumbbells and $\varphi$ is the orientation difference.

M7 

Simulation of $N = 16$ H-like rotors initially connected in a close-loop chain.

M8 

Simulation of 16 X-like rotors. This simulation shows the self-assembly of long-lasting chains with $N_{ch}>3$ without pillars.

M9 

Simulation of 20 X-like rotors initially arranged in a linear interlocked chain.

%In figure \ref{fig:rotation_offset}a, we show a situation corresponding to the repulsive part of the potential for which the rotational activity is going to act as a separating force. The repulsive interaction from the point of view of robot 1 is indeed very steep in the sense that there is very little chance that a fluctuation could help robot 1 to counteract against the rotation and lead to a bound state. However, to reach a bound state from situation figure \ref{fig:rotation_offset}b, it is not required to have large fluctuation that would bring back robot 1 and there is a repetition of attempts for every half turn of the particle.  
%\begin{figure}[h!]
%    \centering
%    \includegraphics[width=\linewidth]{figure_rotation_offset.png}
%    \caption{Representation of (a) a contact configuration and (b) a non-contact configuration.}
%    \label{fig:rotation_offset}
%\end{figure}

%\bibliography{biblio}
%apsrev4-2.bst 2019-01-14 (MD) hand-edited version of apsrev4-1.bst
%Control: key (0)
%Control: author (8) initials jnrlst
%Control: editor formatted (1) identically to author
%Control: production of article title (0) allowed
%Control: page (0) single
%Control: year (1) truncated
%Control: production of eprint (0) enabled
%

\end{document}